%% file: main.tex
\documentclass[letter,reqno]{cts}
\title{
Asymptotic Behavior of Random Time-Inhomogeneous Markovian Quantum Dynamics
}

\usepackage{authblk}

\author[1]{Lubashan Pathirana\thanks{lpk@math.ku.dk}}
\author[2]{Jeffrey Schenker\thanks{schenke6@msu.edu}}
\affil[1]{Department of Mathematical Sciences and QMATH, University of Copenhagen, Denmark.}
\affil[2]{Department of Mathematics, Michigan State University, USA.}
\date{}

\begin{document}

\pagenumbering{arabic}
\lhead{\thepage}
\maketitle
\vspace{-1cm}
\begin{abstract}
     We study the asymptotic behavior of continuous-time, time-inhomogeneous Markovian quantum dynamics in a stationary random environment. 
     Under mild faithfulness and eventually positivity-improving assumptions, the normalized evolution converges almost surely to a stationary family of full-rank states, and the normalized propagators converge almost surely to a rank-one family determined by these states. 
     Beyond a disorder-dependent threshold, these convergences occur at exponential rates that may depend on the disorder; when the environment is ergodic, the rate itself is deterministic.
     When the dynamical propagators display vanishing maximal temporal stochastic correlation, convergence in stochastic expectations for above limits is faster than any power of the time separation,
     and improves to exponential rates when the dynamical propagators display stochastically independent increments. 
     These expectation bounds yield disorder-uniform high-probability estimates.
     The framework does not require complete positivity or trace preservation and encompasses random Lindbladian evolutions and collision-model dynamics.
\end{abstract}
\include*{Sections/1_intro}

\include*{Sections/2_prelim}

\include*{Sections/3_basic_results}
\include*{Sections/4_mixing}
\include*{Sections/5_theorems}
\addcontentsline{toc}{section}{Acknowledgements}
\section*{Acknowledgements}
\noindent Author J.S. acknowledges the support of the National Science Foundation under Grant No. $2153946$. 
Author L.P. would like to thank Albert H. Werner for insightful discussions and thank Villum Fonden for its support with the Grant No. $25452$ and Grant No. $60842$ as well as via the QMATH Centre of Excellence Grant No. $10059$. 
\addcontentsline{toc}{section}{Bibliography}
\printbibliography
\end{document}

%% file: Sections/1_intro.tex
\section{Introduction}\label{section:intro}

The state of a quantum system is described by a \emph{density matrix} $\rho$, a positive semidefinite operator of trace one on the system Hilbert space. 
The dynamical transformation, the evolution, of the system state $\rho_s$ at time $s$ into the state $\rho_t$ at time $t$ is described by a linear propagator $\phi{s}{t}$: $\rho_t=\phi{s}{t}(\rho_s)$.  
For an isolated quantum system, the propagators are unitary transformations on the system Hilbert space. Specifically, $\phi{s}{t}(\rho_s)=U(t,s)\rho_s U(t,s)^\dagger$ where $U(t,s)$ is a unitary map.  
More generally, for an \emph{open quantum system}, the propagators that govern the evolution are a family of \emph{completely positive and trace-preserving} (abbreviated as CPTP) maps $\{\phi{s}{t}\}_{t \geq s}$ acting on bounded linear operators over the system Hilbert space. 
We refer the reader to a standard textbook in quantum information theory, such as \cite{wilde2013quantum, nielsen2010quantum,watrous2018theory}, for mathematical definitions of the terms mentioned above.  

A family of propagators $\{\phi{s}{t}\}_{t \geq s}$ characterizing the dynamics of an open quantum system is called \emph{time-homogeneous Markovian} if $\phi{s}{t}=\tilde\oldphi_{t-s}$ where $\tilde\oldphi_{s}$ is a one-parameter family of maps that satisfies the semigroup property:
\begin{equation}
    \tilde\oldphi_{t_2}\circ \tilde\oldphi_{t_1} 
        = 
            \tilde\oldphi_{t_1+t_2}, \quad \forall t_1, t_2 \ge 0\, .
\end{equation}

Such a family of dynamical maps (with other properties) is called a \emph{quantum dynamical semigroup} \cite{alicki2007quantum}.  
We refer the reader to \cite{rivas2012open, breuer2002theory} for details. 
In general, the dynamics of an open system is not exactly Markovian due to memory effects and back-reaction from the environment.  
However, many systems are physically known to be well approximated by Markovian evolutions on suitable time scales; see, for example, \cite[Chapter 5]{rivas2012open} and references therein. 
    
A norm-continuous quantum dynamical semigroup is known to be generated by a linear map $\mcL$ \cite{lindblad1976generators, gorini1976completely}, called the \emph{generator} of the semigroup so that 
\begin{equation}
\label{eq:master}
         \diff{\oldphi_t(\rho)}{t} 
            = 
                \mcL(\oldphi_t(\rho)), \quad        
         \quad \rho(t_0) = \rho_0\, ,
\end{equation}
with  
\[
    \mcL (\rho) 
        =  -i[\mbH_S, \rho] 
            + \sum_{j} \xi_j 
                \left(
                    V_j\rho V_j\adj - \dfrac{1}{2}\left\{V_j\adj V_j,\rho \right\}
                \right) \, .
\]
where $\mbH_S$ is the system Hamiltonian and $\xi_j$ are nonnegative. 

The operator $\mcL$ is called the Gorini–Kossakowski–Lindblad–Sudarshan (GKLS) generator, or simply the Lindbladian. 
The equation \cref{eq:master} is called a \emph{Master Equation}. 
The most general form of a master equation allows for a time-dependent generator and time dispersion:
    \begin{equation}\label{eq_general_time_dep_master}
        \diff{\rho(t)}{t} 
            = \mcL(t,t_0)(\rho(t)) + \int_{t_0}^t \mcK(s,t_0)(\rho(s))\, \dif s\, , \quad \rho(t_0) = \rho_0 \, .
    \end{equation}
Here, $\mcL(\,\cdot\,,t_0)$ is a time-dependent Lindblad operator, and $\mcK(\,\cdot\,,t_0)$ is a super-operator valued kernel depending on the initial time. 
As indicated, both $\mcL$ and $\mcK$ may explicitly depend on the initial time $t_0$. 
We say that the system is \textit{Markovian} if $\mcK\equiv 0$ and $\mcL(t,t_0)$ is independent of $t_0$, i.e., $\mcL(t,t_0) := \mcL(t)$. 

For a Markovian time-dependent master equation, 
    \begin{equation}\label{eq_time_dep_master}
        \diff{\rho(t)}{t} 
            = \mcL(t)(\rho(t)) , \quad \rho(t_0) = \rho_0 \, ,
    \end{equation}
the solution satisfies
    \begin{equation}\label{eq:propagator_solution}
    \rho(t)=\phi{t_0}{t} (\rho(t_0)) \, ,
    \end{equation}
with the \emph{propagator} $\phi{s}{t}$ given by
    \begin{equation}\label{eq:solution_for_master}
        \phi s t 
            = \mcT\left\{
                \exp\left(\int_{s}^t \mcL(r) \, \dif r\right)
                \right\}\, ,
    \end{equation}
where $\mcT$ denotes the formal time ordering operator.
This family of dynamical propagators satisfies the composition law:
    \begin{equation}\label{eq:the_composition_law}
        \phi{r}{t} 
            = 
                \phi{s}{t}\circ\phi{r}{s} \quad \forall \ r\leq s\leq t \, .
    \end{equation}
Following \cite{rivas2010entanglement}, we shall call an open system \emph{time-inhomogeneous Markovian} if the dynamics is given by \cref{eq:propagator_solution} with the propagators $\{\phi{s}{t}\}_{s\leq t}$ a family of CPTP dynamical maps satisfying  \cref{eq:the_composition_law}.
\begin{remark}\label{remark:markovian}
    The terminology ``Markovian'' stems from the fact that, due to the composition law \cref{eq:the_composition_law}, at any time $s>t_0$ the future states $\{\rho(t)\}_{t>s}$ depend on the past states $\{\rho(t)\}_{t<s}$ only through the present state $\rho_s = \phi{t_0}{s}(\rho(t_0))$. 
    Indeed, the composition law \cref{eq:the_composition_law} resembles the Chapman–Kolmogorov equations for the distribution of solutions to a classical Markov process \textemdash e.g., see \cite{kallenberg1997foundations}. 
    As such, a Markovian open system is said to be \emph{memoryless}.  
    It is important to note that a family of CPTP dynamical maps satisfying \cref{eq:the_composition_law} may nonetheless generate non-Markovian temporal correlations \cite{milz2019completely}. 
    Sometimes \cref{eq:the_composition_law}  is referred to as \emph{CP-divisibility}, to clearly distinguish it from classical Markovianity.
\end{remark}
\subsection{Random Time-inhomogeneous Markovian Dynamics}\label{section:inhomogeneous_markov}
    In this paper, we consider the evolution of a time-inhomogeneous Markov open system with the dynamical propagators given by a stochastic process. 
    Formally, we suppose that $\{\phi{s}{t}\}_{s\leq t}$ consists of random maps
        \[
           \Omega \ni \omega \mapsto  \phi{s}{t}^\omega \in \mcL^{(2)}(\mcH_S) \quad \text{for all } \ s\le t \ , \ s, t \in \mbR \, ,
        \]
    where 
    \begin{enumerate}
        \item $(\Omega,\mcF,\pr)$ is a probability space, which we assume to be a standard Borel space \textemdash see \cref{section:ergodic} for more details, and
        \item $\mcL^{(2)}(\mcH_S)$ denotes the set of ``super-operators'' on the system Hilbert space $\mcH_S$, that is, the vector space of linear homomorphisms of the vector space $\mcL(\mcH_S)$ of linear homomorphisms of $\mcH_S$. 
    \end{enumerate}
    Throughout this article, we make the simplifying assumption that $\mcH_S$ is finite-dimensional and consequently, we identify the space $\mcL(\mcH_S)$ of linear operators on $\mcH_S$ with $\matrices$, the space of complex $D\times D$-dimensional matrices, for some fixed finite $D$.  
    Thus, $\mcL^{(2)}(\mcH_S)$ is isomorphic to $\mbM_{D^2}$, but not canonically so.

    We assume that the family of propagators $\{\phi{s}{t}\}_{s\leq t}$ is \emph{stochastically stationary in time}. Formally, this means that the joint probability distribution (law) of the collection of dynamical maps $\{\phi{s}{t}\}_{s\leq t}$, is identical to that of the time-shifted family $\{\phi{s+h}{t+h}\}_{s\le t}$ for every $h\in \mathbb{R}$.
    To implement this invariance, we assume the existence of a one-parameter group $\theta_h:\Omega \to \Omega$, $h \in \mbR$, of probability-preserving transformations (see \Cref{section:ergodic} below for precise definitions). 
    We shall refer to $\ergsys$ as a \emph{stationary environment}. 
    When this group of transformations, $\{\theta_h\}_{h\in \mbR}$, is also ergodic, we shall call $\ergsys$ an \emph{ergodic environment}.  
    With this terminology, we introduce the following definition:
    
    \begin{dfn}\label{dfn:random_interaction_model}
        A family of positive super-operator valued random maps $\{\phi{s}{t}\}_{s<t}$ defined on a stationary environment $\ergsys$ is called a family of \emph{dynamical propagators in a stationary environment} if
         the following conditions hold with probability $1$:
            \begin{enumerate}
                \item $\phi{s}{t}^\omega \circ \phi{r}{s}^\omega = \phi{r}{t}^\omega$  for all $r\le s\le t$, and
                \item $\phi{s+h}{t+h}^\omega = \phi{s}{t}^{\theta_h(\omega)}$ for $s\le t$ and $h\in \mbR$.
            \end{enumerate}
        If we further assume that $\ergsys$ is ergodic, we shall refer to $\{\phi{s}{t}\}_{s<t}$ as the \emph{dynamical propagators in an ergodic environment}. 
    \end{dfn}

    \begin{remark}
            By a statement of the form ``With probability one, $A_\omega(x)$ for all $x\in R$,'' where $\omega \mapsto A_\omega (x)$ is a random predicate and $R$ is a set, we indicate ``There is an event $E_0\subset \Omega$ of full measure such that for each $\omega \in E_0$ and $x\in R$ the statement $A_\omega(x)$ holds.'' 
            When $R$ is uncountable, as in \Cref{dfn:random_interaction_model}, this is a strictly stronger statement than ``For all $x\in R$, $A_\omega(x)$ with probability one,'' which indicates ``For each $x\in R$ there is a set $E_x$ of full measure such that for all $\omega \in E_x$ the statement $A_\omega(x)$ holds.'' 
        \end{remark}

    We now describe the basic assumptions on the propagators $\{\phi{s}{t}\}_{s\le t}$ that are required for our main results.   
    For further context, we refer the reader to Section~\ref{section:ergodic}.
    Henceforth, we suppress the explicit dependence on the random parameter $\omega$ when there is no risk of confusion. 
    Often $\Omega$ is referred to as the \emph{disorder configuration} and $\omega\in\Omega$ a \emph{disorder realization}. 

    Denote by $\states$ the convex set of density matrices (quantum states of the system $S$),  consisting of all positive semidefinite matrices with unit trace. 
    A super-operator $\varphi \in \mcL^{(2)}(\mcH_S)$ is called \emph{positive} if $\varphi(\rho)\ge 0$ for any $\rho \in \states$. 
    The adjoint $\varphi^\dagger$ of a super-operator $\varphi \in \mcL^{(2)}(\matrices)$ is defined using the Hilbert-Schmidt inner product:
    \begin{equation}\label{eq:adjoint}
       \Tr(A^\dagger \varphi^\dagger(B))  =  \Tr((\varphi(A))^\dagger B) \quad \text{for all } A, B \in \matrices ,
    \end{equation}
    where $A^\dagger$ denotes the conjugate transpose of a matrix $A\in \matrices$.

    \begin{assumption}\label{assumption:1} 
        With probability one, for each $s\le t$ the random map $\phi{s}{t}$ and its adjoint $\phiadj{s}{t}$ are \emph{positive} and \emph{faithful}, that is $\ker{\phi{s}{t}} \cap  \mbS_D = \ker{\phiadj{s}{t}}\cap \mbS_D=\emptyset$.
    \end{assumption}

    \begin{remark} 
        For propagators consisting of CPTP maps $\phi{s}{t}$ obtained from a time-dependent Lindbladian via a master equation, as in \cref{eq:solution_for_master}, \Cref{assumption:1} is automatically true.
        However, as in \cite{movassagh2022ergodic}, neither trace-preservation nor \emph{complete} positivity is required for our results. 
        It is useful to note that if the maps $\phi{s}{t}$ are trace-preserving, then $\ker{\phi{s}{t}} \cap  \mbS_D$ is necessarily empty. 
        However, there are CPTP families satisfying the composition law \cref{eq:the_composition_law} that do not satisfy $\ker{\phiadj{s}{t}}\cap \mbS_D=\emptyset$, e.g., if $\phi{s}{t}(X)=PXP + SXS^\dagger$ for all $s\le t$ with $P$ a non-trivial projection and $S$ an isometry from $\operatorname{ran} P^\perp$ to $\operatorname{ran} P$.
    \end{remark} 
    A super-operator $\oldphi\in \mcL^{(2)}(\matrices)$ is said to be \emph{strictly positive} or \emph{positivity improving} if, for any density matrix, $\rho \in \states$ we have that $\oldphi(\rho)$ is strictly positive definite, i.e., $\oldphi(\rho)$ is positive definite and satisfies $\ker{\oldphi(\rho)}=\{0\}$.  

    Given a family of dynamical propagators $\{\phi{s}{t}\}_{s\le t}$ on the stationary environment $(\Omega,\mathcal{F},\mathbb{P},\{\theta_h\}_{h\in\mathbb{R}})$, we introduce the following random times: 
        \begin{equation}\label{eq:tau+}
        \tau^+(\omega) = \inf\{ t >0 : \phi{0}{t}^\omega \text{ is strictly positive} \} \ , 
        \end{equation}
    and  
        \begin{equation}\label{eq:tau-}
        \tau^-(\omega) = \sup\{ s< 0 : \phi{s}{0}^\omega \text{ is strictly positive} \} \ .
        \end{equation}
    Our results will require variously that $\tau^+<\infty$ with probability one or $\tau^->-\infty$ with probability one. 
    We will show below that $\pr\{\tau^+<\infty\}=\pr\{\tau^->-\infty\}$ \textemdash see \Cref{lemma:tau}.  
    Thus, these amount to the same assumption.
    \begin{assumption}\label{assumption:2} 
        With probability one, $\tau^+ <\infty$ and $\tau^->-\infty$.
    \end{assumption}

    We highlight an important consequence of combining Assumptions~\ref{assumption:1} and~\ref{assumption:2}:
    
    \begin{remark} 
        If the dynamical propagators in a stationary environment satisfy assumptions \ref{assumption:1} and \ref{assumption:2}, it follows that, with probability one, $\phi{0}{t}$ and $\phi{s}{0}$ are strictly positive for all sufficiently large $t$ and $s$. 
        Indeed, the composition law \cref{eq:the_composition_law} and these assumptions together imply that $\phi{0}{t}$ is strictly positive for all $t> \tau^+$ and $\phi{s}{0}$ is strictly positive for all $s< \tau^-$ \textemdash see \cref{prop:strict_positivity_starting_at_0} below. 
    \end{remark}

     Throughout this article, we assume that $\{\phi{s}{t}\}_{s\leq t}$ is a family of dynamical propagators in a stationary environment satisfying the assumptions above. 
     We are interested in the asymptotic behavior of the dynamical maps. 
     As such, this article can be considered as the continuous-parameter counterpart of the setting in \cite{movassagh2022ergodic, movassagh2021theory, pathirana2023law}. 
     However, in contrast, we do not assume that $\ergsys$ is ergodic.  
     As noted above, we do not require the maps to be CPTP, but only that each $\phi{s}{t}$ is a positive map.  
     We do not assume the maps are trace-preserving or are given by solving a master equation as in \eqref{eq:solution_for_master}. 
     Nor do we require that the process $\{\phi s t\}_{s\leq t}$ be sample path continuous, namely that with probability one $(s,t) \mapsto \phi{s}{t}$ is continuous. 

\subsection{Main Results}\label{section:main_results}
    Our main results concern the asymptotic behavior of the dynamical propagators in two regimes: given a fixed time $s_0$, the asymptotic behavior of $\phi{r}{s_0}$ as the initial time $r \to -\infty$ as well as the asymptotic behavior of $\phi{s_0}{t}$ as the final time $t\to\infty$. 
    Since we do not assume the propagators $\phi{s}{t}$ are trace-preserving, their action does not necessarily preserve the set of quantum states of $\mcS$. Accordingly, we consider the \emph{projective action}: $\oldphi\proj\rho=\oldphi(\rho) /\tr{\oldphi(\rho)}$, whenever the denominator is strictly positive. 
    Under Assumption~\ref{assumption:1}, we have $\ker{\phi{s}{t}} \cap \mathbb{S}_D = \emptyset$ and $\ker{\phi{s}{t}^\dagger}\cap \mathbb{S}_D = \emptyset$ with probability one, ensuring that the projective actions of $\phi{s}{t}$ and $\phi{s}{t}^\dagger$ are almost surely well-defined on $\mathbb{S}_D$, for all $s<t$.

    \begin{restatable}[]{thm}{zresult}
    \label{thm:existence_of_z_0}
        Let $\{\phi{s}{t}\}_{s\le t}$ be a family of dynamical propagators in a stationary random environment $\ergsys$,  satisfying Assumptions \ref{assumption:1} and \ref{assumption:2}.
        Then there exist:
        \begin{itemize}
            \item[(i)] two stationary families of random full-rank states $\{Z_t\}_{t\in\mbR},\{Z'_t\}_{t\in\mbR}$, and
            \item[(ii)] a random variable $\mu:\Omega\to (0,1)$, invariant under the time-shifts $\{\theta_h\}_{h\in\mbR}$, 
        \end{itemize}
        such that the following hold almost surely: 
            
            \smallskip
            
        For each $r\ge0$, there exists a $\mbN$-valued random variable $T_{\mu,r}$, a finite random threshold, such that 
        for any state $\rho\in\states$
            \begin{equation}
            \label{eq:existence_of_z_0}
                  \norm{\phi{s}{t}\cdot \rho - Z_t} \le 2\mu^{-r-s-2} \quad \text{ for all } |t| \leq r \text{ and } s < -(T_{\mu,r} + r), 
            \end{equation}
            and
            \begin{equation}
            \label{eq:existence_of_z_0'}
                \norm{\phiadj{s}{t}\cdot \rho - Z_s'}  \le 2\mu^{t-r-2} \quad \text{ for all } |s| \leq r \text{ and }t > T_{\mu, r} + r, .
            \end{equation}
         Furthermore, in the special case where $\ergsys$ is an ergodic environment, we have that $\mu$ is a deterministic constant. 
    \end{restatable}

     Theorem \ref{thm:existence_of_z_0} is one of the fundamental results of the present paper.  
     It shows that the long-time behavior of the dynamics is independent of the initial state $\rho$.  
     This can be viewed as a continuous-time generalization of the discrete-time results in~\cite{movassagh2022ergodic}, extended to strictly stationary random dynamical environments.

    \begin{bigcor}
         With probability $1$, for any initial state $\rho$, we have that 
         \[
            \displaystyle{\lim_{s\to -\infty}}\phi{s}{t}\cdot \rho 
                = Z_t 
                    \quad \text{ for all }\quad t\in \mbR\,
        \]
        and
         \[
            \displaystyle{\lim_{t\to \infty}}\phiadj{s}{t}\cdot \rho 
                = Z_s'
                    \quad \text{ for all } \quad s\in \mbR\, .
        \]
    \end{bigcor}
     
     Since $\norm{\phi{s}{t}\cdot \rho - Z_t}\le 1$ it follows immediately from dominated convergence that we have also convergence in expectation
         \begin{equation}\label{eq:convinexp}
            \lim_{s \to-\infty}\avg{\sup_{\rho \in \states} \sup_{-r\le t\le r} \norm{\phi{s}{t}\cdot \rho - Z_t}} \, ,
         \end{equation}
    with a similar result for $\phiadj{s}{t}$.

    Our next result concerns the \emph{normalized propagators}
    \[
       \widetilde{\phi{s}{t}}\;:=\;
       \frac{\phi{s}{t}}{\Tr\!\bigl[\phiadj{s}{t}(\mbI)\bigr]},
    \]
    obtained by rescaling each $\phi{s}{t}$ with the scalar  
    $\Tr\!\bigl[\phiadj{s}{t}(\mbI)\bigr]>0$ (the positivity follows from Assumption \ref{assumption:1}).  
    We prove that $\widetilde{\phi{s}{t}}$ converges as the time separation becomes large, in \emph{either} direction, to a rank‑one super‑operator determined by the stationary families $\{Z_t\}_{t\in\mbR}$ and $\{Z'_t\}_{t\in\mbR}$ supplied by Theorem \ref{thm:existence_of_z_0}.  Define
    \begin{equation}
    \label{eq:sup_op}
        \opp{s}{t}(\, \cdot \,)  = \tr{Z'_s(\, \cdot \,)}Z_t \,.
    \end{equation}
    Then we have the following result.

    \begin{restatable}[Convergence of normalized propagators to rank-one limits]{thm}{supopthm}
    \label{thm:super-op-convergence}
        Let $\{\phi s t\}_{s\le t}$ be a family of dynamical propagators in a stationary random environment $\ergsys$ satisfying Assumptions~\ref{assumption:1} and~\ref{assumption:2}. 
        Let $\mu\in(0,1)$ and the stationary families $\{Z_t\}_{t\in\mathbb{R}}$, $\{Z'_t\}_{t\in\mathbb{R}}$ be as in \Cref{thm:existence_of_z_0}, and define
            \[
              \opp{s}{t}(X):=\Tr\!\bigl[Z'_s X\bigr]\;Z_t .
            \]
        Then with probability $1$, for each $r\ge 0$,
            \[
              \norm{\frac{\phi{s}{t}}{\Tr[\phiadj{s}{t}(\mbI)]}-\opp{s}{t}}
              \le 8\,\mu^{-\,r-s-2}
              \quad\text{for all } |t|\le r \text{ and } s<-(T_{\mu,r}+r),
            \]
        and
            \[
              \norm{\frac{\phi{s}{t}}{\Tr[\phiadj{s}{t}(\mbI)]}-\opp{s}{t}}
              \le 8\,\mu^{\,t-r-2}
              \quad\text{for all } |s|\le r \text{ and } t> T_{\mu,r}+r ,
            \]
        where $T_{\mu,r}$ is the same random threshold in \Cref{thm:existence_of_z_0}. 
    \end{restatable}

    The following is the immediate consequence of \Cref{thm:super-op-convergence}.
    
    \begin{bigcor}
            In particular, with probability one, we have that 
            \[
                \displaystyle\lim_{s\to -\infty} 
                \norm{\dfrac{\phi s t}{\tr{\phiadj s t(\mbI)}} - \opp{s}{t}} 
                = 
                0 \quad \text{ for all } \quad  t\in \mbR\, ,
            \]
            and
            \[
                \displaystyle\lim_{t\to\infty}
                \norm{\dfrac{\phi s t}{\tr{\phiadj s t(\mbI)}} - \opp{s}{t}} 
                = 0
                \quad \text{ for all } \quad  s \in \mbR\,.
            \]
    \end{bigcor}

\subsubsection{Convergence in Expectation and Deviation Bounds}

    \Cref{thm:existence_of_z_0} and \Cref{thm:super-op-convergence} establish almost sure convergence with rates: i.e. convergence rates and prefactors that are dependent on the random realization (i.e. disorder variable, $\omega\in\Omega$).
    In many random operator settings, this is the typical scenario: one obtains detailed asymptotic behavior for almost all realizations, yet uniform bounds---those valid for \emph{all} realizations simultaneously---are often unattainable without further assumptions on the distribution. 
    However, suppose one assumes that the \emph{maximal temporal stochastic correlation} between the dynamical propagators vanishes asymptotically. In that case, as we shall see below, we can obtain high probability uniform rates. 
    
    To do this, one needs the convergence rate for expectations. 
    The result below achieves this and states that if one has that the discrete sub-process $\{\phi {n} {n+1}\}_{n\in\mbN_0}$ has asymptotically vanishing maximal temporal stochastic correlation in the sense that the sub-process $\{\phi {n} {n+1}\}_{n\in\mbN_0}$ is $\rho$-mixing (see \cref{section:mixing} for proper definition) then the convergence of the stochastic expectation is super-polynomial---or even exponential in the case the dynamical propagators have independent increments.

    \begin{restatable}[]{thm}{fourthresult} \label{thm:mixing}
        Let $\{\phi s t\}_{s<t}$ be a family of dynamical propagators defined on a stationary environment satisfying the assumptions \ref{assumption:1} and \ref{assumption:2}. 
        Let $\{Z'_t\}_{t\in\mbR}$ and $\{Z_t\}_{t\in\mbR}$ be the family of states obtained in \Cref{thm:existence_of_z_0}, and let $\{\opp{s}{t}\}$ be the rank-one super-operators from \Cref{thm:super-op-convergence}. 
        Suppose that the discrete sub-process $\{\phi {n} {n+1}\}_{n\in\mbN_0}$ is $\rho$-mixing (or $\oldpsi$ or $\oldphi$-mixing),  then for any $p\in\mbN$ we have that,
            \begin{enumerate}
                % \item $\avg{\norm{\phiadj s t \proj \updelta - Z'_s}} \lesssim_p  
                %         \dfrac{1}{(t-s-1)^p}$, for all $\updelta\in\states$
                \item $\avg{\norm{\phi s t \proj \updelta - Z_t}} \lesssim_p  
                        \dfrac{1}{(t-s)^p}$, for all $\updelta\in\states$ and for all $t > s+1$. 
                \item $\avg{ \norm{\dfrac{\phi s t}{\tr{\phiadj s t(\mbI)}} - \opp{s}{t}}} \lesssim_p \dfrac{1}{(t-s)^p}$ for all $t > s+1$. 
            \end{enumerate}
        Furthermore, if the family $\{\phi {n} {n+1}\}_{n\in\mbN_0}$, is jointly independent, then the above convergence can be improved from super-polynomial to exponential, i.e.
        \[\avg{\ldots} \lesssim e^{-\delta(t-s)}\, ,\]
        for some constant $\delta>0$. 
    \end{restatable}
         
    With the obtained rates for the expectation, one is now able to obtain rates that are uniform in the disorder parameter, with high probability. 

    \begin{restatable}[]{bigcor}{probresult}
    \label{thm:large_probability}

    Let $\{\phi s t\}_{s<t}$ be a family of dynamical propagators (see \cref{dfn:random_interaction_model}) satisfying the assumptions \ref{assumption:1} and \ref{assumption:2}.  If the discrete sub-process $\{\phi {n-1} n\}_{n\in\mbN}$ is $\rho$-mixing(or the $\oldpsi$-mixing or $\varphi$-mixing), then for a given $p\in\mbN$ there is a constant $C'_p >0$ (that depends only on $p$) such that 
        \begin{equation}
            \pr\left\{
            \norm{\phi{s}{t}\cdot \updelta - Z_t} \leq (t-s)^{-p} \right\} \geq 1 - C'_p (t-s)^{-p}\, ,
        \end{equation}
    for all $\updelta\in\states$ and
        \begin{equation}
          \pr\left\{ \norm{\dfrac{\phi s t}{\tr{\phiadj s t (\mbI) }}- \Xi_{s,t}} \leq (t-s)^{-p} \right\} \geq 1 - C'_p (t-s)^{-p}\, ,
        \end{equation}
    for any $t>s+1$.
    In addition, if the family of dynamical propagators $\{\phi {n-1} n\}_{n\in\mbN}$  is jointly independent, then we have that there is a constant $C>0$ and $\gamma> 0$ such that for all $s<t$ with $t-s>0$ we have that, 
        \begin{equation}
            \pr\left\{
            \norm{\phi{s}{t}\cdot \updelta - Z_t} \leq e^{-\gamma(t-s)} \right\} \geq 1 - C e^{-\gamma(t-s)}\, ,
        \end{equation}
    for all $\updelta\in\states$ and
        \begin{equation}
        \pr\left\{ \norm{\dfrac{\phi s t}{\tr{\phiadj s t (\mbI) }}- \Xi_{s,t}} \leq e^{-\gamma(t-s)} \right\} \geq 1 - C e^{-\gamma(t-s)} \, .
        \end{equation}
    \end{restatable}
\subsection{Related Literature in Other Random Dynamics}\label{section:related_work}
        As noted previously, the results presented here extend those of \cite{movassagh2021theory,movassagh2022ergodic} from a discrete-time setting to the continuous-time case. Furthermore, because we do not assume ergodicity of the stationary environment, $\ergsys$, our work also generalizes the discrete-time results in \cite{movassagh2021theory,movassagh2022ergodic} to the strictly stationary case. 

        Discrete-time evolution is often motivated by repeated indirect measurements, often referred to as \emph{repeated-interaction dynamics} or \emph{collision models} for an open quantum system. Several authors have studied random discrete \-time dynamics, modeling these repeated interactions with a stochastic environment; see, for instance,  \cite{bruneau2008random, bruneau2010infinite, nechita2012random, bougron2022markovian}. For the deterministic case, there are notable works bridging discrete and continuous time dynamics \cite{pellegrini2008existence, bauer2013repeated, attal2006repeated}. 

        Recall that a (time-independent) GKLS generator (or \emph{Lindbladian}) is generally split into a Hermitian (Hamiltonian) part $\mcL_U :=-i[\mbH_S, \rho]$ and a dissipative part $\mcL_D := \sum_{j} \xi_j (V_j\rho V_j\adj - \frac{1}{2}\{V_j\adj V_j,\rho\}) $. Here  $V_j$ are called the \emph{jump operators}, and each of the quantities $\xi_k$ are called \emph{rate variables}. 
        Random Lindbladians have been studied in various contexts, for example by sampling the Hamiltonian from the Gaussian unitary ensemble (GUE) or the jump operators from the Ginibre ensemble; see \cite{denisov2019universal,can2019random} (and \cite{onorati2017mixing} for random unitary evolution). In some treatments, the rate parameters $\xi_j$ are also drawn from specific probability distributions \cite{budini2005random}.

%% file: Sections/2_prelim.tex
\section{Preliminaries, Notation and Definitions}
     We denote by $\mbR^+$ (resp. $\mbR^-$) the set of positive (resp. negative) real numbers. We denote by $\mbN$ the positive integers and by $\mbN_0 = \mbN\cup\{0\}$. 
     We denote by $\mbZ^-$ the set of negative integers and by $\mbZ^-_{0} = \mbZ^{-}\cup\{0\}$. For $x\in\mbR$, $\floor{x}$ denotes the greatest integer less than or equal to $x$ and $\ceil{x}$ denotes the smallest integer greater than or equal to $x$. 
     We shall use $x \lesssim y$ as a shorthand to denote the inequality $x \leq c y$ where $c$ is some positive constant. We will also use the notation  $x \lesssim_p y$ to indicate $x\leq c_p \, y$ where $c_p$ is a $p$-dependent positive constant. It need not be the case that each use of $\lesssim$ (or $\lesssim_p$) indicates the same constant $c$ (or $c_p$). 
    \subsection{Probability Theory and Ergodic Flows}\label{section:ergodic}
        Let  $(\Omega,\mcF,\pr)$ be a probability space, where $\Omega$ is the underlying set, $\mcF$ is the $\sigma$-algebra of $\pr$-measurable subsets of $\Omega$ and $\pr$ is the associated probability measure. 
        A function $\theta:\Omega\to\Omega$ is called a \emph{probability-preserving transformation} if $\theta$ is $\mcF$-measurable and for any event $E\in\mcF$ one has $\pr[\theta^{-1}(E)] = \pr(E)$. 
        Given a group $G$, a family $\{\theta_g\}_{g\in G}$ of such transformations is called a \emph{group of probability-preserving transformations}  if 
            \begin{enumerate}
            	\item For all $g_1, g_2 \in G$, $\theta_{g_1} \circ \theta_{g_2} = \theta_{g_1 \circ g_2}$, where $\circ$ denotes the group operator on $G$, and
            	\item $\theta_e(\omega) =\omega$ for all $\omega \in \Omega$, where $e$ denotes the identity element of $G$.
            \end{enumerate}
        When $G=\mbR$ with the group operation $\circ=+$ given by addition, the family is called a \emph{one-parameter group of probability-preserving transformations.}

        Given a group of probability-preserving transformations $\{\theta_g\}_{g\in G}$, we see that $\theta_g$ is a bijection for each $g\in G$.  
        Indeed, we must have $\theta_g \circ \theta_{g^{-1}} = \theta_e = \theta_{g^{-1}} \circ \theta_g$. As such, $\theta_g$ is invertible with the inverse $\theta_{g^{-1}}$. A probability space  $(\Omega,\mcF,\pr)$, together with a group of probability-preserving transformations,  $\{\theta_g\}_{g\in G}$ is referred to as a \emph{probability-preserving system}.  
        When the group is a one-parameter group, we refer to the system $\ergsys$ as a \emph{stationary environment}, as above. 

        Events $A,B \in \mcF$ are said to be \emph{essentially equal}, written $A \overset{a.s.}= B$, if $\pr(A\triangle B)=0$, where $\triangle$ denotes the symmetric difference of sets.
        A family of probability-preserving transformations $\{\theta_g\}_{g\in G}$ on a probability space $(\Omega,\mcF,\pr)$, with $G$ any set (not necessarily a group), is called \emph{ergodic in the weak sense} if any event $E\in \mcF$ that is essentially $\theta_g$-invariant for all $g\in G$, has probability $0$ or $1$.  Here an event $E\in\mcF$ is called \emph{essentially $\theta_g$-invariant} if $\theta_g^{-1}(E) \overset{a.s.}= E$.
        Such a family of probability-preserving transformations is sometimes referred to as a \emph{weakly ergodic family}. 
        The family $\{\theta_g\}_{g\in G}$ is called \emph{ergodic in the strong sense} if any event $E\in \mcF$ that is $\theta_g$-invariant for all $g\in G$ has probability either $0$ or $1$, where an event $E\in\mcF$ is called $\theta_g$-invariant if $\theta_g^{-1}(E) =E$. 
        It is clear from these definitions that a weakly ergodic family is also strongly ergodic. 

        It is well known \cite{halmos_operator_1942} that when the indexing set is countable $G = \mathbb{N}$ or $G=\mathbb{Z}$, the two versions of ergodicity are equivalent for any probability space $(\Omega,\mcF,\pr)$. Furthermore, by a result of Mackey \cite[Theorem 3]{mackey1962point} (also see \cite{Moore_1966}), if $(\Omega, \mcF, \pr)$ is a standard Borel probability space and $G$ is a locally compact separable group, then any strongly ergodic group of probability-preserving transformations $\{\theta_g\}_{g\in G}$ on $(\Omega,\mcF,\pr)$ is, in fact, weakly ergodic. 
        Here, a probability space $(\Omega,\mcF,\pr)$ is called a \emph{standard Borel probability space} if the underlying measurable space $(\Omega,\mcF)$ is the Borel space associated to a separable, completely metrizable topological space (a \emph{Polish space}).  
        As such, throughout this article, we assume that $(\Omega,\mcF,\pr)$ is a standard Borel space equipped with a one-parameter group of probability-preserving transformations $\{\theta_h\}_{h\in \mbR}$. As above, we call the system $\ergsys$ an \emph{ergodic environment} if the family $\{\theta_h\}_{h\in\mbR}$ is an ergodic family. 
        
        In an ergodic environment $\ergsys$, a key consequence of ergodicity is that any function that is essentially $\theta_h$-invariant for all $h\in \mbR$ is almost surely a constant.  
        Here, a function $f$ is called essentially $\theta_h$-invariant if $\pr\{f\circ\theta_h \neq f\} = 0$. We refer the reader to \cite{brin2002introduction, walters2000introduction} for an in-depth treatment of measure-preserving transformations and flows. In what follows, we will write that a function (or set) is $\{\theta_h\}_{h\in\mbR}$-essentially invariant to mean that the function (or set) is $\theta_h$-essentially invariant for all $h\in \mbR$. 

        Given a predicate $A_\omega$ depending on $\omega\in\Omega$, we use the terminology ``$A$ \emph{holds with probability $1$},'' abbreviated as ``$A$ holds w.p. $1$,'' to indicate that there is an event $E$ with $\Pr(E)=1$ such that $A_\omega$ is true for $\omega \in E$. 
        In other words, the set $\{\omega : A_\omega \text{ does not hold }\}$ is a subset of an event of measure $0$  (a \emph{sub-null} set).  
        We denote the measurability of a function $f$ with respect to a $\sigma$-algebra $\mcA$ by the abbreviated notation $f\in \mcA$. The average of a function $f$, $\int f(\omega) \ d\pr(\omega)$ is denoted by $\mbE_{\pr}[f]$, or by $\avg{f}$ if the underlying probability measure is clear.  
        Whenever a random variable $f$ takes values that are themselves maps (a matrix, super-operator etc.), we denote it by $f^\omega(\,\cdot\,)$ or by $f_\omega(\, \cdot \,)$.

        With the notations above it is useful to point out that the propagators defined in \cref{dfn:random_interaction_model} can be modeled through a \emph{random dynamical system}: We have a metric dynamical system $\ergsys$ and a measurable mapping $\tilde{\oldphi} : \mbR\times\Omega\times\states\to\states $ where $\tilde\oldphi(s,\omega, x)= \phi 0 s(\omega)(x)$ for all $x\in\states$ with the (perfect) cocycle property $\tilde\oldphi(t+s,\omega) = \tilde\oldphi(t, \theta_s(\omega))\circ\tilde\oldphi(s,\omega)$ for all $s,t\in\mbR$ and $\omega\in\Omega$ see \cite{arnoldRDS, arnold1995random} for details.  Also, the composition law \cref{eq:the_composition_law}, which now is satisfied almost surely (if one restricts to discrete time) resembles a relation of transition kernels of a (classical) Markov chain in a stationary environment i.e. a Markov chain with random transition kernels on a random stationary environment \cite{cogburn1984ergodic, cogburn1980markov}.

    \subsection{Quantum States, Super-operators and Contraction Coefficient of a Positive Super-operator}\label{section:q_operations}
        Recall that $\mbM_D$ denotes the space of $D\times D$ complex matrices.  
        We denote by $\mapspace$ the space of linear operators from $\mbM_D$ to $\mbM_D$, called \emph{super-operators}. We consider the space $\matrices$ with the trace norm $ \norm{M} := \norm{M}_1 = \tr{(M\adj M)^{1/2}}$
        and for super-operators $\oldphi\in\mapspace$ we use the operator norm generated by the trace norm on $\matrices$: 
            \begin{equation}\label{eq:operatornorm}
                \norm{\oldphi} := \norm{\oldphi}_{1\to1} = \sup\{\norm{\oldphi(X)}_1 : \norm{X}_1\leq 1\} \, . 
            \end{equation}
        We denote by $\inner{\,\cdot\,}{\,\cdot\,}$ the Hilbert-Schmidt inner product on matrices, 
            \begin{equation}
                \inner{A}{B} = \tr{A^\dagger B}\, .
            \end{equation}
        We refer the reader to any standard textbook on matrix analysis, such as \cite{horn2012matrix, bhatia2013matrix}, for some of the standard matrix facts mentioned above.
        For $\oldphi\in\mapspace$ we can find the adjoint operator of $\oldphi$, denoted by $\oldphi\adj$, which is the unique operator in $\mapspace$ that satisfies
            \begin{equation}
                \inner{\oldphi(A)}{B} = \inner{A}{\oldphi\adj(B)}
            \end{equation}
        for all $A,B\in\matrices$. 
        
        As described above, $\states$ denotes the convex set of positive semidefinite matrices of trace $1$.  We denote by $\statesint$ the interior of $\states$, that is, 
            \begin{equation}
                \statesint = \left \{ A \in \states : \ker{A}=\{ 0\} \right \} \, ,
            \end{equation}
        and by $\partial \states$ the boundary $\partial \states = \states\setminus\statesint$.  A density matrix $\rho \in \partial \states$ has at least one $0$ eigenvalue.  
        
        A super-operator $\oldphi\in\mapspace$ is called \emph{positive} if $\oldphi(X)\ge$ for all $X\ge 0$ and is called \emph{strictly positive} if $\oldphi(X)>0$ for all non-zero positive semidefinite $X\ge 0$.  
        The following collection of super-operators is at the center of the present work:
            \begin{equation}\label{eq:posandkernel}
            \dpspace = \left \{ \oldphi \in \mapspace : \oldphi \text{ is positive and } \ker{\oldphi}\cap\states =\emptyset \right \} \ .
            \end{equation}
        Note that the maps $\phi{s}{t}$ of a dynamical propagator satisfying \Cref{assumption:1} satisfy $\phi{s}{t}\in \dpspace$ and $\phi{s}{t}^\dagger\in \dpspace$. Given $\oldphi \in \dpspace$, we define the \emph{projective action} on $\states$ via
            \begin{equation}
            \label{eq:projective}
                \oldphi\proj  \rho = \frac{1}{\tr{\oldphi(\rho)}} \oldphi(\rho) \ .
            \end{equation}
        \begin{definition}[Contraction coefficient]
            Let $\oldphi\in \dpspace$.  The \emph{contraction coefficient of $\oldphi$}, denoted $\cnum{\oldphi}$, is defined as follows: 
            \begin{equation}\label{eq:cnum}
                \cnum{\oldphi} 
                    = 
                        \sup \{\dis{\oldphi\proj A}{\oldphi\proj B} : A,B \in \states \} \, ,
            \end{equation}
            where $\dis{\,\cdot\,}{\,\cdot\,}$ is the following metric on $\states$:
            \begin{equation}\label{eq:d}
        		\dis{A}{B} \ 
                        := \ 
                            \frac{1 - m(A,B)m(B,A)}{1+m(A,B)m(B,A)} \, ,
        	\end{equation}
            where $m(A,B)  =  \sup \{\lambda\ge 0 : \lambda B\leq A\}$.
        \end{definition}

        \begin{remark} The metric $\dis{\,\cdot\,}{\,\cdot\,}$ and the contraction coefficient $\cnum{\cdot}$ were defined in \cite{movassagh2022ergodic}, adapting notions defined in \cite{hennion1997limit} for positive matrices and vectors. 
        \end{remark} 
        We recall here some useful properties of the metric $\mathrm{d}$:
        \begin{prop}[{\cite[Lemma 3.9]{movassagh2022ergodic} and \cite[Lemma 4.3]{raquepas2024}}]\label{prop:d}
            Let $\rho,\updelta \in \mbS_D$, then we have that 
            \begin{enumerate}
                \item $\frac{1}{2}\norm{\rho-\updelta} \leq \dis{\rho}{\updelta} \leq \frac{1}{\eta(\rho)}\norm{\rho-\updelta} $, with $\eta(\rho)$ the smallest eigenvalue of $\rho$.
                \item  $\sup_{\rho,\updelta \in \states} \{\dis{\rho}{\updelta}\} = 1$. 
                \item If $\rho \in \statesint$ and $\updelta\in\states$ then $\dis{\rho}{\updelta} = 1$ if and only if $\updelta \in \partial \states$.
            \item   On $\statesint$ the trace-norm topology and the $\dis{\cdot}{\cdot}$-induced topology are homeomorphic.
            \end{enumerate}
        \end{prop}
        \begin{remark} 
            On the interior of $\statesint$, convergence with respect to the metric $\mathrm{d}$ is equivalent to convergence in the trace norm. However, $\mathrm{d}$ separates $\statesint$ from $\partial \states$ in a very strong way.
            The metric space $(\states,\mathrm{d})$ is non-compact, and has uncountably many connected components (one of which is $\statesint$). 
        \end{remark}
        We now turn to properties of the contraction coefficient.
        \begin{lemma}\label{lemma:cnum}
            If $\oldphi \in \dpspace$, then
                \begin{enumerate} 
                    \item $\dis{\oldphi\proj \rho}{\oldphi\proj \updelta} \leq \cnum{\oldphi} \dis{\rho}{\updelta}$  for all $\rho,\updelta \in \states$.
                    \item  $\cnum{\oldphi} \leq 1$ and if $\oldphi$ is strictly positive then $\cnum{\oldphi}< 1$.
                    \item If there exist $\rho,\updelta$ such that $\oldphi\proj \rho \in \statesint$ and $\oldphi\proj \updelta \in \partial \states$, then $\cnum{\oldphi} = 1$.
                    \item If $\oldpsi\in \dpspace$, then $\cnum{\oldphi\circ\oldpsi} \leq \cnum{\oldphi}\cnum{\oldpsi}$.
                    \item If furthermore $\oldphi^\dagger\in \dpspace$, then  $\cnum{\oldphi} = \cnum{\oldphi\adj}$, and
                    \begin{equation}\label{eq:coverinterior}
                        \cnum{\oldphi} = \sup_{\updelta,\updelta' \in \statesint} \dis{\oldphi \proj \updelta}{\oldphi\proj\updelta'}\, .
                    \end{equation}
                \end{enumerate}
        \end{lemma}

            \begin{proof}
            All of these items except for \cref{eq:coverinterior} are stated in \cite[Lemma 3.10]{movassagh2022ergodic}.  To prove \cref{eq:coverinterior}, we note that by the second-to-last inequality in the proof of \cite[Lemma 3.10]{movassagh2022ergodic}, we have for any $\rho,\rho'\in \states$
            $$ m(\oldphi\proj \rho, \oldphi\proj \rho') m(\oldphi\proj \rho',\oldphi\proj\rho) \ge \inf_{\updelta,\updelta'\in \statesint} m(\oldphi^\dagger\proj \updelta,\oldphi^\dagger\proj \updelta') m(\oldphi^\dagger\proj \updelta',\oldphi^\dagger\proj \updelta) \ ,$$
            from which it follows that
            $$ \dis{\oldphi\proj \rho}{ \oldphi\proj \rho'} \le \sup_{\updelta,\updelta'\in \statesint} \dis{\oldphi^{\dagger}\proj \updelta}{\oldphi^\dagger\proj \updelta'}\le \cnum{\oldphi^\dagger} = \cnum{\oldphi} \, .$$
            Taking the supremum of $\rho,\rho'\in \states$ leads to the desired result.
            \end{proof}
        \begin{cor}
        \label{cor:c_iff_s_p}
            An immediate consequence of the lemma above is that, if $\oldphi,\oldphi\adj \in \dpspace$ then $\cnum{\oldphi}<1$ if and only if $\oldphi$ is strictly positive.
        \end{cor}
        \begin{proof}
            One of the directions is simply item $2$ of \cref{lemma:cnum}. 
            For the converse if $\ker{\oldphi\adj}\cap\states = \emptyset$ then we have that $\oldphi\proj\statesint\subseteq \statesint$. 
            To see this we first establish that $\oldphi\proj (\mbI/D) = \oldphi\proj\mbI \in \statesint$: 
            if not, let $P$ be the projection onto the kernel of $\oldphi\proj \mbI$, then $0 = \tr{P\oldphi(\mbI)} = \tr{\oldphi\adj(P) \mbI}$ giving us that $P/\norm{P} \in \ker{\oldphi\adj}\cap\states$, which is a contradiction. 
            The fact that $\oldphi \proj \statesint\subseteq \statesint$ now follows from the simple observation that for any $X\in\statesint$ there is a $\delta>0$ such that $X\ge\delta\mbI$. 
            Having established that whenever $\oldphi\adj \in \mcP(\matrices)$ we have that  $\oldphi \proj \statesint\subseteq \statesint$, we use item 3 of \cref{lemma:cnum} to finish the claim. 
            Suppose that $\oldphi$ is not strictly positive, thus there is some $X\in\states$ such that $\oldphi \proj X \in \partial\states$ but as $\oldphi\proj \statesint\subset\statesint$ we also have that there is some $Y\in\states$ such that $\oldphi \proj Y \in \statesint$, yielding that $\cnum{\oldphi}=1$ from item $3$ of \cref{lemma:cnum}.
        \end{proof}
        
        The following lemma about the continuity of $\oldpsi \mapsto \cnum{\oldpsi}$ is useful in particular for establishing the measurability of $\cnum{\oldpsi}$ when $\oldpsi$ is a $\dpspace\cap \dpspace^\dagger$-valued random map, where $\dpspace^\dagger$ denotes the set of positive linear operators in $\mapspace$ with $\ker{\oldphi\adj} \cap \states = \emptyset$.

        \begin{lemma} \label{lemma:continuityOfC} The contraction coefficient 
        	$\cnum{\,\cdot\,}$ is lower semi-continuous on $\dpspace\cap \dpspace^\dagger$.
        \end{lemma}

            \begin{proof}
                To begin we note that if $\oldphi\in \dpspace\cap\dpspace^\dagger$ and $\updelta\in \statesint$ then $\oldphi\proj\updelta\in \statesint$, i.e., $\oldphi\proj\statesint\subset \statesint$.  Indeed, since $\oldphi^\dagger\in \dpspace$ and $\updelta \in \statesint$, we conclude that $\inner{\rho}{\oldphi(\updelta)} = \inner{\oldphi^\dagger (\rho)}{\updelta} > 0 $ for any $\rho\in \states$.  It follows that $\ker{\oldphi(\updelta)}=\{0\}$ so that $\oldphi \proj \updelta \in \statesint$ as claimed.
                
                Now let $(\oldphi_n)_{n\in\mbN}$ be a sequence in $\dpspace\cap \dpspace^\dagger$ and suppose that $\oldphi_n \to \oldphi\in \dpspace\cap \dpspace^\dagger$, with convergence in operator norm. Since $\rho \mapsto \Tr{\oldphi(\rho)}$ is continuous and positive on the compact set $\states$, we conclude that there is $\eta>0$ with $\Tr{\oldphi(\rho)}\ge \eta$ for all $\rho\in \states$. Then, by the triangle inequality,
                \begin{equation}\label{eq:phiconv} \norm{\oldphi_n\proj\rho - \oldphi\proj\rho}  \leq  \frac{1}{\tr{\oldphi(\rho)}} \Bigl ( \norm{ \oldphi_n(\rho)-\oldphi(\rho)}  +  \abs{\tr{\oldphi_n(\rho)} - \tr{\oldphi(\rho)}} \Bigr )   \le  \frac{2}{\eta} \norm{\oldphi_n-\oldphi} \, ,\end{equation}
                for all $\rho \in \states$. It follows that $\norm{\oldphi_n\proj\rho - \oldphi\proj\rho} \to 0$ uniformly in $\rho\in \states$, as $n\to\infty$.  

                Now let $\updelta,\updelta'\in \statesint$.  Then
                
                $$\dis{\oldphi\proj\updelta}{\oldphi\proj\updelta'} \leq \dis{\oldphi\proj\updelta}{\oldphi_n\proj\updelta} + \dis{\oldphi_n\proj\updelta'}{\oldphi\proj\updelta'} + \cnum{\oldphi_n} \, .$$  
                Since $\oldphi \proj \updelta$, $\oldphi \proj \updelta'\in \statesint$, we conclude from \cref{eq:phiconv} and part 1 of \Cref{prop:d} that
                $$\dis{\oldphi\proj\updelta}{\oldphi\proj\updelta'} \le \liminf_{n\to \infty} \, \cnum{\oldphi_n} \ .$$
                Taking the supremum over $\updelta,\updelta'\in \statesint$, we find that $\cnum{\oldphi}\le \liminf_n \cnum{\oldphi_n}$, from which the result follows.
            \end{proof}

\subsection{Eventual Strict Positivity for Quantum Dynamics in a Stationary Environment}\label{section:random_dynamics}
    In this section, we consider the random times $\tau^\pm$ beyond which the propagator $\phi{s}{t}$ is strictly positive\textemdash see \cref{eq:tau+} and \cref{eq:tau-}.  In general, we do not expect that $\tau^+=-\tau^-$, however, it turns out that the events on which these times are finite are essentially the same. 

    First, note that from \Cref{lemma:cnum} and \Cref{assumption:1} we have that $\cnum{\phi r t} \leq \cnum{\phi s t}\cnum{\phi r s}$ and that $\cnum{\oldphi_{s+h,t+h}^\omega} = \cnum{\oldphi_{s,t}^{\theta_h\omega}}$  for all $r< s< t$ and $h\in\mbR$ with probability $1$. Now we present the following proposition. 
    
    \begin{prop}
    \label{prop:strict_positivity_starting_at_0}
        With probability $1$,  $\phi{s}{0}$ and $\phi{0}{t}$ are strictly positive for all $s<\tau^-$ and $t>\tau^+$, with $\tau^\pm$ as in \cref{eq:tau+} and \cref{eq:tau-}.
    \end{prop}
        \begin{proof}
            The key observation here is that if $\oldphi$ is strictly positive and $\oldpsi \in \dpspace$, then $\oldphi\circ\oldpsi$ is strictly positive. Indeed, for any $\rho \in \states$ we then have $\oldpsi(\rho)\neq 0$ from which we conclude that $(\oldphi\circ \oldpsi) \proj\rho = \oldphi \proj (\oldpsi(\rho))\in \statesint$ since $\oldphi$ is strictly positive and $\oldpsi\in\dpspace$.
            
            Let $\omega$ be in the event that $\tau^+<\infty$ and $\tau^->-\infty$ and \cref{assumption:1} holds.  Consider first $s<\tau^-$. Then, by the definition of $\tau^-$ we are able to find $s' \in (s,\tau^-)$ such that $\phi{s'}{0}$ is strictly positive. By the composition law $\phi{s}{0} = \phi{s'}{0}\circ\phi{s}{s'}$, which is strictly positive by the above observation.

            Turning now to the strict positivity of $\phi{0}{t}$ for $t>\tau^+$, we first note that a super-operator $\oldphi\in \mcL(\matrices)$ is strictly positive if and only if $\oldphi^\dagger$ is strictly positive. Indeed, $\oldphi$ is strictly positive if and only if $\inner{\updelta}{\oldphi(\rho)}>0$ for any $\updelta, \rho\in \states$.  Since $\inner{\updelta}{\oldphi(\rho)}= \inner{\rho}{\oldphi^\dagger(\updelta)}$ the claim follows. Thus, it suffices to show that $\phiadj{0}{t}$ is strictly positive. Now fix $t'\in (\tau^+,t)$ such that $\phi{0}{t'}$ is strictly positive. Then $\phi{0}{t}=\phi{t'}{t}\circ\phi{0}{t'}$ and thus $\phiadj{0}{t}=\phiadj{0}{t'}\circ \phiadj{t'}{t}$, so that $\phiadj{0}{t}$ is strictly positive by the argument above.
        \end{proof}

    \begin{lemma}
    \label{lemma:tau}
        If $\{\phi{s}{t}\}_{s\le t}$ is a dynamical propagator in a stationary random environment that satisfies \Cref{assumption:1}, then $\{\tau^+<\infty\} \overset{a.s}{=} \{\tau^->-\infty\}$.
    \end{lemma}

        \begin{proof} 
            Let $S^+= \{\omega : \tau^+(\omega) <\infty\}$. Note that $\phi{0}{t}$ is strictly positive if and only if $\cnum{\phi{0}{t}} < 1$ for some $t$.  Since $\cnum{\phi{0}{\ceil{t}}}\le \cnum{\phi{0}{{t}}}$, we find from \Cref{prop:strict_positivity_starting_at_0} that 
            \[
                S^+  =  \bigcup_{n\in \mbN} \{ \omega : \cnum{\phi{0}{n}^\omega} < 1 \} \, .
            \]
            Similarly, with $S^-=  \{\omega : \tau^-(\omega) >-\infty\}$, we have
            $$ S^- = \bigcup_{n\in \mbN} \{ \omega : \cnum{\phi{-n}{0}^\omega} < 1 \} \, ,$$
          
            Since
            $$ \{ \omega : \cnum{\phi{-n}{0}^\omega} < 1 \} = \theta_n( \{ \omega : \cnum{\phi{0}{n}^\omega} < 1 \} ) \subset \theta_n(S^+) \overset{a.s}{=} S^+$$
            for each $n\in\mbN$, we conclude that $\pr(S^-\setminus S^+)=0$.  By a similar argument, reversing the roles of $S^+$ and $S^-$, we conclude that $\pr(S^+\setminus S^-)=0$.  Thus $S^+\overset{a.s}= S^-$ as claimed.
        \end{proof}

%% file: Sections/3_basic_results.tex
\section{Proofs of \texorpdfstring{\Cref{thm:existence_of_z_0}}{Theorem 1} and \texorpdfstring{\Cref{thm:super-op-convergence}}{Theorem 3}}
\label{section:forward}
    In the sections below, we shall always restrict our attention to the event of full probability where \Cref{assumption:1} and \Cref{assumption:2} hold. 

\subsection{Contraction properties of the dynamical propagators} 

    \begin{lemma}
    \label{lemma:kappa}
        There exists a non-negative random variable $\kappa$ such that, with probability one, $\kappa <1$ and
        \begin{equation}\label{eq:kappa}
         \ln\kappa 
                    = 
                        \lim_{t\to\infty} \dfrac{1}{t} \, \ln{\cnum{\phi{0}{t}}} = \lim_{s\to -\infty} \dfrac{1}{|s|}\ln{\cnum{\phi{s}{0}}} \, .
        \end{equation}  
    \end{lemma}
        \begin{proof}
            The proof is based on Kingman's subadditive ergodic theorem \cite{kingman1973subadditive, kingmanSubadditiveProcesses1976}.  
            Consider the discrete parameter sub-family $\{\phi{m}{n}:m,n\in\mbZ, \, m<n\}$. Observe that with probability $1$, for all $p,q,r\in\mbZ$ with $p<q<r$ one has  
            \begin{equation}\label{eq:discrete-subad}
                \ln{\cnum{\phi{p}{r}}} \leq \ln{\cnum{\phi{q}{r}}} + \ln{\cnum{\phi{p}{q}}} \, .
                \end{equation}
            Furthermore, we have that 
            \begin{equation}\label{eq:discrete-invariant}
                    (\ln{\cnum{\phi{m}{n}}})_{m<n}\overset{d}{=} (\ln{\cnum{\phi{m+1}{n+1}}})_{m<n} \ ,
            \end{equation}
            where $\overset{d}=$ denotes equality in (joint) distribution. Therefore, by Kingman's subadditive ergodic theorem, the limits 
                \begin{equation}\label{limit:over_n}
                    \xi^+ := \lim_{n \to \infty} \dfrac{\ln{\cnum{\phi{0}{n}}}}{n} \quad \text{and} \quad \xi^- := \lim_{n\to \infty} \dfrac{\ln{\cnum{\phi{-n}{0}}}}{n}
                \end{equation}
            exist with probability one and define $\{\theta_{n}\}_{n\in \mbZ}$-essentially invariant random variables. Furthermore, we also have that 
            
                \begin{equation}
                    \xi^+ \overset{\text{a.s.}}
                    =\lim_{n\to\infty} \dfrac{\mbE[\ln\cnum{\phi{0}{n}} | \mcS_{\mbZ} ] }{n} \overset{\text{a.s.}}
                    = \inf_{n\in\mbN}\dfrac{\mbE[\ln\cnum{\phi{0}{n}} | \mcS_{\mbZ} ]}{n}
                \end{equation}
            and similarly for $\xi^-$,
                \begin{equation}
                    \xi^- \overset{\text{a.s.}}
                    =\lim_{n\to\infty} \dfrac{\mbE[\ln\cnum{\phi{-n}{0}} | \mcS_{\mbZ} ]}{n}  \overset{\text{a.s.}}
                    = \inf_{n\in\mbN}\dfrac{\mbE[\ln\cnum{\phi{-n}{0}} | \mcS_{\mbZ} ]}{n}\, .
                \end{equation}
            Here $\mcS_{\mbZ}$ denote the $\sigma$-algebra of $\{\theta_{n}\}_{n\in \mbZ}$-essentially invariant sets.

            Because of the $\theta$-invariance of the conditional expectations $\mbE[\ln\cnum{\phi{-n}{0}} | \mcS_{\mbZ} ]$, we have that $\xi^+ = \xi^-$ almost surely. 
            Furthermore as $\cnum{\, \cdot \,} \leq 1$ we also have that $\xi^+ = \xi^- \le 0$. But from \Cref{assumption:2} (cf. \Cref{lemma:tau}) we have that a.s.\ for $\omega$ the existence of some $n\in\mbN$ such that $\ln\cnum{\oldphi^\omega_{0,n}} < 0$. As such $\xi^-=\xi^+ < 0$

            We now define $\ln\kappa = \xi^+$, i.e. $\kappa = e^{\xi^+}$, whence $\kappa \in [0,1)$ almost surely.
%%%%%%%%%%%%%%%%%%%%%%%%%%
%%%%%%%%%%%%%%%%%%%%%%%%%%
%%%%%%%%%%%%%%%%%%%%%%%%%%

            It remains to show that \cref{eq:kappa} holds for the continuous limits $t\to\infty$ and $s\to-\infty$. To this end,  note that, with probability one, we have for all $t\in\mbR^+$ that 
                \begin{equation}\label{ineq:lemma_kappa_1}
                    \ln{\cnum{\phi{0}{t}}} 
                        \leq 
                            \ln{\cnum{\phi{n}{t}}} + \ln{\cnum{\phi{0}{n}}} 
                        \leq 
                            \ln{\cnum{\phi{0}{n}}}
                \end{equation}
            and
                \begin{equation}\label{ineq:lemma_kappa_2}
                    \ln{\cnum{\phi{0}{n+1}}} 
                        \leq 
                            \ln{\cnum{\phi{t}{n+1}}} + \ln{\cnum{\phi{0}{t}}} 
                        \leq 
                            \ln{\cnum{\phi{0}{t}}} \ ,
                \end{equation}
            where $n = \floor{t}$.
            For both inequalities \eqref{ineq:lemma_kappa_1} and \eqref{ineq:lemma_kappa_2} we have used that $\ln{\cnum{\oldpsi}}\leq0$ for $\oldpsi\in\dpspace$. It follows that
                \begin{equation}
                    \frac{\ln{\cnum{\phi{0}{n+1}}}}{n+1} \cdot \frac{n+1}{t} 
                        \leq 
                            \frac{\ln{\cnum{\phi{0}{t}}}}{t} 
                        \leq 
                            \frac{\ln{\cnum{\phi{0}{n}}} }{n} \cdot \frac{n}{t}\, .
                \end{equation}
            Taking the limit $t\to\infty$ we find that $\ln\kappa = \lim_t \frac{1}{t} \ln{\cnum{\phi{0}{t}}}$ with probability one. By a similar argument $\ln\kappa = \lim_t \frac{1}{t} \ln{\cnum{\phi{-t}{0}}} $ almost surely.
        \end{proof}

     As in the discrete parameter case \cite{movassagh2022ergodic}, if $\ergsys$ is an ergodic system (i.e., if the one-parameter group $\{\theta_h\}_{h\in\mbR}$ is ergodic), then $\kappa$ is deterministic. 

    \begin{lemma}\label{lemma:kappa_is_invariant}
        The random variable $\kappa$ is essentially $\{\theta_h\}_{h\in\mbR}$-invariant. If $\ergsys$ is an ergodic system, it follows that $\kappa$ takes a deterministic value in $[0,1)$ with probability one.
    \end{lemma}

        \begin{proof} 
        We will show that $ \kappa\circ\theta_h - \kappa=0$ almost surely for each $h\in \mbR$. To this end, first consider $h>0$ and observe that, for $t>0$,
            \[ \frac{1}{t+h} \ln{\cnum{\phi{0}{t+h}^\omega}}  \leq \dfrac{t}{t+h}
                                \left(\frac{1}{t}\ln{\cnum{\phi{0}{t}^{\theta_h(\omega)}}}
                                    +
                                    \frac{1}{t} \ln{\cnum{\phi{0}{h}^\omega}}
                                \right) \ , \]
        by \Cref{lemma:cnum} and stationarity. Taking the limit $t\to\infty$ and using \Cref{lemma:kappa}, we find that $\kappa \leq \kappa\circ\theta_h$ almost surely. Since $0\le \kappa \le 1$, it follows that $0\le \kappa\circ\theta_h - \kappa \le 1$.  Averaging and using shift invariance, we conclude that $\kappa\circ\theta_h -\kappa$ is a non-negative random variable with mean zero, and thus that $\kappa\circ\theta_h-\kappa=0$ almost surely. For $h<0$, we have for $t>|h|$ that
        \[
                \frac{1}{t} \ln{\cnum{\phi{0}{t}^{\theta_{h}(\omega)}}}
                        = 
                            \frac{1}{t}
                                \ln{\cnum{\phi{h}{t+h}}}
                        \leq 
                            \frac{1}{t}
                                \ln{\cnum{\phi{h}{0}}}
                            + 
                            \frac{
                                t+h
                                }{t}
                            \cdot \frac{1}{t+h}
                                \ln{\cnum{\phi{0}{t+h}}} \, .
          \]
            Taking the limit $t\to\infty$ and using \Cref{lemma:kappa}, we obtain that $\kappa \circ\theta_h -\kappa \le 0$ almost surely. A similar argument to the above establishes that $\kappa\circ \theta_h -\kappa \ge 0$ almost surely.    
        \end{proof}

        By the essential shift invariance of $\kappa$, we have $\frac{1}{t}\ln{\cnum{\phi{s}{s+t}}} \to \kappa$ as $t\to \infty $ for any $s$, and similarly for $\frac{1}{t}\ln{\cnum{\phi{s-t}{s}}}$.  We can use this observation to show that $\cnum{\phi{s}{s+t}}$ and $\cnum{\phi{s-t}{s}}$ decay exponentially fast to $0$ as $t\to \infty$, with uniform estimates for $s$ in a bounded interval.

    \begin{lemma}\label{lemma:cnum_to_0}
        Let $\lambda \in (0,1)$ and define $\mu_\lambda(\omega) = \lambda \kappa(\omega) + (1-\lambda)$. 
        With probability $1$ we have that for each $r\in\mbN_0$ 
        \begin{equation}
        \label{eq:cnum_to_0}
            \sup_{-r\le s\le r} \max \left \{ \cnum{\phi{s}{s+t}} \, , \cnum{\phi{s-t}{s}} \right \} \leq \mu_\lambda^{t -2} \quad \text{ for all }t \ge T_{\lambda,r}\, ,\end{equation}
        where $T_{\lambda,r}$ is an almost surely $\mbN$-valued random variable, for each $r\in\mbN$. 
        In particular, with probability one, for each real-valued $r\ge 0$ we have that inequality \eqref{eq:cnum_to_0} holds
        for $T_{\lambda,r} = T_{\lambda,\ceil{r}}$.
    \end{lemma}
%%%%%%%%%%%%%%%%%%%%%%%%%%
%%%%%%%%%%%%%%%%%%%%%%%%%%
%%%%%%%%%%%%%%%%%%%%%%%%%%
        \begin{proof}
            Because of invariance we have that the discrete limit $\cnum{\phi{m}{m+n}^\omega}^{1/n} \to \kappa(\omega)$ with probability $1$ for each $m\in\mbZ$. 
            Since $\mu_\lambda(\omega)> \kappa(\omega)$ a.s., we have that 
            $$ N_m^+(\omega) := \min \{ n\in\mbN : \cnum{\phi{m}{m+p}^\omega}^{1/p} \le \mu_{\lambda}(\omega) \text{ for all } p\in\mbN \text{ with } p \ge n\}$$
            defines an $\mbN$-valued random variable such that $\cnum{\phi{m}{n+m}^\omega}\le \mu_\lambda(\omega)^n$ for all $n \ge N_m^+(\omega)$.  The restriction to a discrete limit, guarantees that $N_m^+(\omega)$ is measurable:
            \[ 
                \{\omega : N_m^+(\omega) \le n \} = \bigcup_{k=1}^n \bigcap_{\substack{p\in \mbN\\p\ge k}} \{ \omega :\cnum{\phi{m}{m+p}^\omega}^{1/p} \le \mu_{\lambda}(\omega)\} \, . 
            \]
            Similarly, using \Cref{lemma:kappa}, we can find an a.s.\ $\mbN$-valued random variable $N_m^-(\omega)$ such that, with probability one, $\cnum{\phi{-n+m}{m}}\le \mu_\lambda(\omega)^n$ for all $n\in\mbN$ with $n\ge N_m^-(\omega)$. 
            
            \medskip

            Now given $r\ge 0$, we define 
            \begin{equation}
            \label{eq:T_r}
                 T_{\lambda,r} = \max_{-\ceil{r}\le m \le \ceil{r}} \max \{N_m^+,N_m^-\} + 1 .
            \end{equation}
             With the definitions of $N_m^+$ and $N_m^-$ we have that $T_{\lambda,r}$ is an  $\mbN$-valued random variable and $T_{\lambda,r}\geq 2$ almost surely. 
             Now take any real-valued $t\ge T_{\lambda , r}$ and any real valued $|s|\le r$.
             Firstly, as $t\geq T_{\lambda , r}$ we must have that $\floor{t} \ge N^+_{\ceil{s}}+1$ where we have used that $T_{\lambda,r}$ is integer-valued and that $|s|\leq r$ . Then we also have that $s\le \ceil{s} <\ceil{s} +\floor{t}-1 \le s+t$. Therefore, we have that 

             \[
                \cnum{\phi{s}{s+t}} \leq \cnum{\phi{\ceil{s}}{\ceil{s}+\floor{t}-1}} \leq \mu_\lambda^{\floor{t}-1} \le \mu_\lambda^{t-2}\, .
             \]

            Similarly, we also have that for $t\geq T_{\lambda}$ and $|s|\leq r$ it must be the case that $s-t\leq \floor{s}-\ceil{t}+1 \leq \floor{s} \leq s$. Thus, $\ceil{t}-1 \ge N^-_{\floor{s}}$ and hence,

            \[
                \cnum{\phi{s-t}{s}} \leq \cnum{\phi{\floor{s}-\ceil{t} + 1}{\floor{s}}} \leq \mu_\lambda^{\ceil{t}-1}\leq \mu_\lambda^{t-2}\, .
            \]
            The two inequalities above prove the first part of the lemma and yield inequality \eqref{eq:cnum_to_0}.

            As such, we have obtained that for each $r\ge0$ there is an event with probability $1$ upon which inequality \eqref{eq:cnum_to_0} holds. In particular, for each $r\in\mbN_0$, there is an event with probability $1$ such that the inequality \eqref{eq:cnum_to_0} holds there. Taking the intersection of (countably many) such events, we have that with probability $1$ for each $r\in\mbN$, the inequality \eqref{eq:cnum_to_0} holds. 

            Now consider the event of full probability whose existence was proved above, and let $r\ge0$ be an arbitrary non-negative real number. Then we have that 
            \[
                 \sup_{|s|\leq r} \ \max\{\cnum{\phi{s}{s+t}} ,\cnum{\phi{s-t}{s}}\} \leq  \sup_{|s|\leq \ceil{r}} \max\{\cnum{\phi{s}{s+t}} ,\cnum{\phi{s-t}{s}}\} \le \mu^{t-2} \quad \forall t\geq T_{\lambda, \ceil{r}}
            \]
            on the full probability event.
            But, on the event considered before we have that 
                \[
                    \sup_{|s|\leq \ceil{r}} \ \max\{\cnum{\phi{s}{s+t}} ,\cnum{\phi{s-t}{s}}\} \leq \mu^{t-2} \quad \forall t\geq T_{\lambda, \ceil{r}}\, .
                \]
            The proof is concluded by noting that, by definition, $T_{\lambda, \ceil{r}} = T_{\lambda, r}$.
               \end{proof}

   Convergence of the expectations  $\cnum{\phi{s-t}{s+t}}$ as $t\to \infty$ follows immediately from the previous result:

    \begin{lemma}\label{lemma:expectation_of_cnum_to_0}
        For any $r\ge0$,
        \begin{equation}
            \lim_{t\to\infty} \avg{\sup_{|s| \le r} \ \max\{\cnum{\phi{s}{s+t}},  \cnum{\phi{s-t}{s}}\}} = 0\, .
        \end{equation}
    \end{lemma}
        \begin{proof}
        Since $\sup_{-r\le s\le r} \cnum{\phi{s-t}{s+t}}\le 1$, this follows from \Cref{lemma:cnum_to_0} by dominated convergence.    
        \end{proof}

\subsection{Perron-Frobenius Eigenmatrices}
    Any strictly positive super-operator $\oldpsi\in\mapspace$ is irreducible in the sense of \cite{evans1977spectral}, and therefore has unique right and left Perron-Frobenius eigenmatrices $R,L\in \states$ such that
    $$ \oldpsi(R)= \lambda R \quad \text{and} \quad \oldpsi^\dagger(L) = \lambda L \ , $$
    where $\lambda$ is the spectral radius of $\oldpsi$ (see \cite[Theorem 2.3]{evans1977spectral}). The eigenmatrices $R$ and $L$ are unique up to scaling; we have normalized them by insisting that $\tr{R}=\tr{L}=1$. 

    For the dynamical propagator, we define random matrices $R_{s,t}(\omega)$ and $L_{s,t}(\omega)$ as follows

    \begin{enumerate}
        \item On the event $\{\omega : \cnum{\phi{s}{t}^\omega} < 1 \}$, we define  $R_{s,t}(\omega)$ (resp. $L_{s,t}(\omega)$) to be the unique right (resp. left) Perron-Frobenius eigenmatrix of $\phi{s}{t}^\omega$ in $\states$, as above.
        \item On the event $\{\omega : \cnum{\phi{s}{t}^\omega} = 1\}$, we take $R_{s,t}(\omega)=L_{s,t}(\omega)=\frac{1}{D} \mbI$.
    \end{enumerate}
    Recall that $\oldpsi\in\dpspace$ is strictly positive if and only if $\cnum{\oldpsi}<1$, by \Cref{lemma:cnum}.  It follows that $\omega \mapsto R_{s,t}(\omega)$ and $\omega \mapsto L_{s,t}(\omega)$ are $\states$-valued measurable maps and that, on the event that $\cnum{\phi{s}{t}}<1$, we have

    \begin{equation}\label{eq:L_and_R}
                \phi{s}{t} (R_{s,t}) 
                    = 
                        \Lambda_{s,t}R_{s,t} \quad \text{ and } \quad \phiadj{s}{t}(L_{s,t}) 
                    = 
                        \Lambda_{s,t}L_{s,t}\, ,
    \end{equation}
    with $\Lambda_{s,t}$ the spectral radius of $\phi{s}{t}$.  For our purposes, the definition of $R_{s,t}$ and $L_{s,t}$ on the event that $\cnum{\phi{s}{t}}=1$ is arbitrary; we take $R_{s,t}=L_{s,t}=\frac{1}{D}\mbI$ on this set to avoid issues of measurability in making a choice of Perron-Frobenius eigematrices when they are potentially degenerate.

    \begin{lemma}
    \label{lemma:L_and_R}
        Let $\lambda \in (0,1)$. Then, with probability $1$, we have that for any $r>0$, 
        and for any $|s|\leq r$ and for all $t\geq T_{\lambda, r}$, $\cnum{\phi{s}{s+t}}$ and $\cnum{\phi{s-t}{s}}$ are both strictly less than $1$. In particular, with probability one, \cref{eq:L_and_R} holds as $s\to -\infty$ for $t$ fixed and as $t\to\infty$ for $s$ fixed.
    \end{lemma}

        \begin{proof}
            This follows directly from \Cref{lemma:cnum_to_0} with the fact that $\mu_\lambda<1$ with probability $1$ whenever $\lambda\in(0,1)$ together with the fact that when the family of dynamical propagators satisfy \Cref{assumption:1}, $\cnum{\phi{s}{t}}<1$ if and only if $\phi{s}{t}$ is strictly positive from \Cref{cor:c_iff_s_p}. 
        \end{proof}
\subsection{Proof of Theorem \ref{thm:existence_of_z_0}}

    We now consider asymptotic limits $\lim_{s\to-\infty}R_{s,t}$ and $\lim_{t\to\infty}L_{s,t}$.
    
    \begin{lemma}\label{lemma:z_0}
        Let $\{\phi{s}{t}\}_{s<t}$ be a dynamical propagator in a stationary environment that satisfies  \Cref{assumption:1} and \Cref{assumption:2}. Then
        \begin{enumerate}
            \item For each $t$, the limit 
            \begin{equation}\label{limit:lim_l}
                \lim_{s\to-\infty} R_{s,t} =: Z_t
            \end{equation}
            exists $\pr$-almost surely, and defines a $\statesint$-valued random variable. 
            \item For each $s$, the limit 
            \begin{equation}\label{limit:lim_r}
                \lim_{t\to\infty} L_{s,t} =: Z_s'
            \end{equation}
            exists $\pr$-almost surely, and defines a $\statesint$-valued  random variable. 
        \end{enumerate}
         Furthermore, we have $Z_{t+h}=Z_t\circ \theta_h$ and $Z_{s+h}'=Z_s'\circ\theta_h$ for all $s$, $t$ and $h$. 
    \end{lemma}
        \begin{proof}
            We proceed similarly to the discrete case in \cite{movassagh2022ergodic}. Fix $t\in \mbR$.   Since $\phi{s-h}{t}\cdot \states \subset \phi{s}{t}\cdot \states$ for $h>0$, the family $\{\phi{s}{t} \cdot \states\}_{s\le t}$ is decreasing as $s\to -\infty$.  Each of these sets is compact in the trace norm topology (being a closed subset of $\states$), so the intersection $\cap_{s\le t} \phi{s}{t}\cdot \states $  is non-empty. 
            
            By \Cref{lemma:cnum_to_0}, for $s\le t - T_{\lambda,|t|}$ we have $\cnum{\phi{s}{t}}<\mu_\lambda^{t-s-2}$.  Thus as $s\to -\infty$ we have $\phi{s}{t}\proj\states \subseteq \statesint$ and the diameter of this set converges to zero in the $\dis{\cdot}{\cdot}$ metric. It follows that $\cap_{s\le t} \phi{s}{t}\cdot \states $ consists of a unique element $Z_t\in \statesint$.  Since $R_{s,t} \in \phi{s}{t}\cdot \states$ as $s\to -\infty$ we see that $Z_t=\lim_{s\to-\infty} R_{s,t}$.  Noting that $Z_t = \lim_{n} R_{-n,t}$ as $n\to \infty$ in $\mbN$ we see that $\omega \mapsto Z_t(\omega)$ is a measurable map.  Finally, the invariance $Z_{t+h}=Z_t\circ \theta_h$ follows directly from the stationarity property $\phi{s+h}{t+h}^\omega=\phi{s}{t}^{\theta_h(\omega)}$.

            The proof of the limit \cref{limit:lim_r} and the invariance property of $Z_s'$ is similar.  It is based on the containment $\phiadj{s}{t+h}\cdot \states \subset \phiadj{s}{t}\cdot \states$ since $\phiadj{s}{t+h}=\phiadj{s}{t}\circ \phiadj{t}{t+h}$.  Since $\cnum{\phiadj{s}{t}}=\cnum{\phi{s}{t}}$, \Cref{lemma:cnum_to_0} continues to provide the necessary contraction bound.
        \end{proof}

    Throughout the rest of this section, fix $\lambda\in(0,1)$ and write $\mu:= \mu_\lambda=(1-\lambda) + \lambda\kappa \in (0,1)$. \Cref{lemma:z_0} now  provides an immediate proof of \cref{thm:existence_of_z_0}.

    \zresult*

        \begin{proof}
            The proof of \Cref{lemma:z_0} establishes that 
                \[
                    \sup_{\rho \in \states}\norm{\phi{s}{t}\cdot\rho - Z_t} \leq 2\dis{\phi{s}{t}\cdot\rho }{Z_t} \leq 2\cnum{\phi{s}{t}} \ ,
                \]
           where we have used \Cref{prop:d}. By \Cref{lemma:cnum_to_0} we have that with probability $1$, for each $r>0$
            \begin{equation}
                \sup_{\rho\in\states} \ 
                \sup_{|t|\leq r} \ \norm{\phi{t-a}{t}\cdot \rho - Z_t} \leq 2 \sup_{|t|\leq r}\cnum{\phi{t-a}{t}} \le 2\mu^{a-2}
            \end{equation}
            for all $a \ge T_{\lambda,r}$. In particular for $s \le -r - T_{\lambda,r} \le t - T_{\lambda,r}$ we obtain that
            \begin{equation}
            \label{ineq:exp_decay}
                \sup_{\rho\in \states} \ \sup_{|t|\le r} \norm{\phi{s}{t}\cdot\rho - Z_t}  
                \le 2\mu_\lambda^{-r-s-2} 
            \end{equation} 
            Since above holds with probability $1$ irrespective of $r$ or $s$, we have obtained \cref{eq:existence_of_z_0}, with probability one.
            The proof of \cref{eq:existence_of_z_0'} is similar.
            Finally, taking respective limits in \cref{eq:existence_of_z_0} and \cref{eq:existence_of_z_0'} we obtain that $\displaystyle{\lim_{s\to -\infty}}\phi{s}{t}\cdot \rho = Z_t$ for all $t\in \mbR$, and
            $\displaystyle{\lim_{t\to \infty}}\phiadj{s}{t}\cdot \rho = Z_s'$ for all $s\in \mbR$, concluding the proof.
    \end{proof}

%%%%%%%%%%%%%%%%%%%%%%%%

        We present below an immediate consequence of \Cref{thm:existence_of_z_0}. 
        
        \begin{cor}
        \label{cor:cocycles}
            With probability $1$ , the following hold: 
            \begin{enumerate}
                \item $\phi s t \proj Z_s = Z_t$ for all $s < t$, and 
                \item $\phiadj s t \proj Z'_t = Z'_s$ for all $s < t$. 
            \end{enumerate}
        \end{cor}
        \begin{proof}
            From \Cref{thm:existence_of_z_0} we have that almost surely, for $t\in\mbR$,  it must be the case that for any $\rho\in\states$
            \[
             \lim_{s\to -\infty} \phi{s}{t} \proj \rho = Z_t 
            \]
            Now for fixed $s'<t$ we have that
            \[
                Z_t 
                = 
                \lim_{s\to -\infty} \phi{s}{t} \proj \rho 
                = 
                \lim_{s\to -\infty} (\phi{s'}{t}\circ \phi{s}{s'}) \proj \rho 
                = 
                 (\phi{s'}{t})\proj \lim_{s\to -\infty} \phi{s}{s'} \proj \rho 
             = \phi{s'}{t}\proj Z_{s'} 
            \]
            proof of the second claim in \Cref{cor:cocycles} is similar. 
        \end{proof}

\subsection{Proof of \texorpdfstring{\Cref{thm:super-op-convergence}}{Theorem on rank-one approximation}}

We start with establishing the following lemma

    \begin{lemma}\label{lemma:sup_op_c_bounds}
        With probability $1$ for any $s<t$
         $$\sup_{\rho\in\states}\norm{\dfrac{\phi s t (\rho)}{\tr{\phiadj s t (\mbI)}} - \opp{s}{t}(\rho) } \leq 4\cnum{\phi s t}\, .$$
    \end{lemma}
        \begin{proof}
            From \Cref{cor:cocycles} we have that $\phi s t \proj Z_s = Z_t$ with probability $1$ for all real-valued $s<t$. Thus, we have from \Cref{lemma:z_0} that 
            \begin{equation}\label{ineq:sup_op_lemma_1}
                2\cnum{\phi s t} \geq  \norm{\phi s t \proj \rho - Z_t} = \norm{\phi s t \proj \rho - \phi s t \proj  Z_s}\, ,
            \end{equation}
            for any $\rho\in\states$. Similarly, we also have that for each $\rho\in\states$,
            \begin{equation}
                2\cnum{\phi s t} \geq \norm{\phiadj s t \proj \rho -  Z'_s} \geq \left|\tr{\phiadj s t \proj \rho } - \tr{Z'_s}\right|\, .
            \end{equation}
            In particular, we obtain that 
                \[
                    \norm{\dfrac{\phiadj s t (\mbI)}{\tr{\phiadj s t (\mbI)}} - Z'_s} \leq 2\cnum{\phiadj s t}\, .
                \]
            Thus, for real-valued $s<t$ we have that 
                \begin{equation}\label{ineq:sup_op_lemma_2}
                    \abs{
                    \dfrac
                    {\tr{\phi s t (\rho)}}{\tr{\phiadj s t (\mbI)}}                        - \tr{Z'_s\rho}
                    }
                    \leq \abs{
                    \tr{
                        \dfrac{\phiadj s t (\mbI)\rho}{\tr{\phiadj s t (\mbI)}}
                        }
                         - \tr{Z'_s\rho}
                    }
                        \leq 2\cnum{\phi s t} \, .
                \end{equation}
            Whence we get 
                \begin{equation}\label{ineq:sup_op_lemma_3}
                    \norm{
                    \dfrac
                    {\tr{\phi s t (\rho)}}{\tr{\phiadj s t (\mbI)}} Z_t
                     - \tr{Z'_s\rho}Z_t
                    }
                    \leq 2\cnum{\phi s t} \, .
                \end{equation}
            Now we multiply \ref{ineq:sup_op_lemma_1} by $\alpha = \tr{\phi s t (\rho)} / \tr{\phiadj s t (\mbI)}$ and combining with \ref{ineq:sup_op_lemma_3} to obtain that
                \begin{align}
                    2\cnum{\phi s t }\times \left(1 +\alpha \right) 
                        &\geq  
                            \norm{\dfrac{\phi{s}{t}(\rho)}{\tr{\phiadj s t (\mbI)}} - Z_t\times \frac{\tr{\phi s t (\rho)}}{ \tr{\phiadj s t (\mbI)}}} 
                        + 
                            \norm{\frac{\tr{\phi s t (\rho)}}{ \tr{\phiadj s t (\mbI)}} \times Z_t - \tr{Z'_s\rho}Z_t} \\
                        &\geq
                            \norm{
                            \dfrac{\phi{s}{t}(\rho)}{\tr{\phiadj s t (\mbI)}} -  \tr{Z'_s\rho}Z_t
                            } \, .
                \end{align}
            Therefore each $(s,t)\in\mbR^2$ with $s<t$ we have that
                \begin{equation}\label{ineq:rank_one_approx_cbound}
                    \norm{\dfrac{\phi s t (\rho)}{\tr{\phiadj s t (\mbI)}} - \opp{s}{t}(\rho) } \leq 2\cnum{\phi s t} \times\left(1 + \dfrac{\tr{\phi s t (\rho)}}{\tr{\phiadj s t (\mbI)}}\right) \leq 4\cnum{\phi s t} \, .
                \end{equation}
            Where we have used that $\tr{\phi s t (\rho)} = \tr{\phiadj s t (\mbI) \rho} \leq \tr{\phiadj s t (\mbI)}$, as $\norm{\rho} = 1$.
        \end{proof}
        
    Now we are ready to prove \Cref{thm:super-op-convergence}, which we shall re-state below.
    \supopthm*
    \begin{proof}
        By \Cref{lemma:sup_op_c_bounds}, we have that 
        \[
         \norm{\dfrac{\phi s t}{\tr{\phiadj s t (\mbI)}} - \opp s t } \leq 8\cnum{\phi s t}\, .
        \]
        This is because any $X \in \matrices$ can be written as $X = \sum_{i=1}^4 \alpha_i \rho_i$ where each $\rho_i\in\states$ and $\sum_{i} |\alpha_i| \leq 2\norm{X}$.  Now from \Cref{lemma:cnum_to_0} we conclude that, w.p. $1$, for each $r>0$
        \[
         \norm{\dfrac{\phi s t}{\tr{\phiadj s t (\mbI)}} - \opp s t } \leq 8 \mu_\lambda^{t-s-2} \, ,
        \]
        for all $|s|\leq r$ with $t-s \geq T_{\lambda,\ceil{r}}$ or for any $|t|\leq r$ with $t-s \geq T_{\lambda, \ceil{r}}$ where $T_{\lambda,r}$ is defined as in the proof of \Cref{lemma:cnum_to_0}. Thus, taking the limit $s \to -\infty$ or $t \to \infty$, we obtained the desired result as claimed. 
    \end{proof}

%% file: Sections/4_mixing.tex
\section{
Mixing Conditions and Proof of \texorpdfstring{\Cref{thm:mixing}}{probability close to one}
}
\label{section:mixing}

\subsection{Mixing coefficients}
    In this section, we introduce the mixing coefficients that enable us to control the decay of stochastic expectations, which in turn yield uniform high-probability bounds.  
    As is clear from \Cref{lemma:sup_op_c_bounds}, controlling the stochastic correlations between (functions of) the dynamical propagators is the key step.
    
    First, we introduce the following notions for random variables. 
    For a random variable $X$ on a probability space, we define the \emph{variance} of $X$ as
        \[
            \vari(X) = \mbE[X^2] - (\mbE[X])^2 \, .
        \]
    For two random elements, $X$ and $Y$ we define the \emph{covariance} of the two random variables by 
        \[
            \cov(X,Y) = \mbE[(X-\mbE[X])(Y-\mbE[Y])] = \avg{XY}-\avg{X}\avg{Y}\, ,
        \]
    and the correlation by 
        \[
            \corr(X,Y) 
                = 
                \begin{cases} 
                    \dfrac{\cov(X,Y)}{\sqrt{\vari(X)}\sqrt{\vari(Y)}}, 
                        & \quad\text{ if } \vari(X), \vari(Y) \neq 0 \ ,\\
                    0 
                        &\quad\text{ if } \vari(X)=0 \text{ or } \vari(Y) = 0\, . 
                \end{cases}
        \]   
    We denote by, $\mathrm{sd}(\, \cdot \,)$, the standard deviation, i.e. the square root
    of the variance.
    
    \begin{dfn}
        Let $(\Omega, \mcF,\pr)$ be a probability space and let $\mcA$, $\mcB$ be two sub $\sigma$-algebras. Then we define the following measures of the correlation between $\mcA$ and $\mcB$:
        \begin{align*}
            \rho(\mcA,\mcB) \,  
                &= \,  \sup \left\{ \left| \corr(X,Y) \right| \ : \ Y\in L^2(\mcA), \  X\in L^2(\mcB), \ X, Y\neq0 \right\}\, ,\\
            \oldpsi(\mcA,\mcB) \, 
                &= \, \sup\left\{\left|1 - \dfrac{\pr(A\cap B)}{\pr(A)\pr(B)} \right|: A \in \mcA, \ B\in\mcB, \ \pr(A),\pr(B)\neq 0\right\}\, , \\
            \varphi(\mcA,\mcB) \, 
                &= \, \sup\left\{\abs{\pr(B | A) - \pr(B)}\ : \ A\in\mcA,\  B\in\mcB,  \ \pr(A)>0\right\}\, .
        \end{align*}
    \end{dfn}
    
        Here $L^p(\mcA)$ for $1\le p< \infty$ denotes the space of $\mcA$-measurable functions $f$ such that $|f|^p$ is integrable. 
        We note that the three conditions $\rho(\mcA,\mcB)=0$, $\oldpsi(\mcA,\mcB)=0$, and $\varphi(\mcA,\mcB)=0 $ are  each equivalent to independence of the two $\sigma$-algebras $\mcA$ and $\mcB$.  
        The quantity $\rho$ is called the \emph{maximal correlation} between $\mcA$ and $\mcB$. 
        We also refer the reader to \cite{bradley2005basic} for a survey on different mixing coefficients, or to \cite{bradley2007introduction} for a detailed treatment of these mixing coefficients. 
        It is evident from the definitions that one has $0\leq \rho(\mcA,\mcB) \leq 1$,  $0\leq \oldpsi(\mcA,\mcB) \leq \infty$, and $0\leq \varphi(\mcA,\mcB) \leq 1$ for any two sub $\sigma$-algebras $\mcA$ and $\mcB$. 

        Out of the three mixing coefficients discussed above, it is straightforward to verify that $\rho$ and $\oldpsi$ are symmetric. However, $\varphi$ is not symmetric. A key use of these mixing coefficients is to get covariance bounds for random variables, as displayed in the lemma below

        \begin{lemma}[{\cite[\S1.2 Theorem 3]{doukhan2012mixing}},  {\cite[Theorem 3.9]{bradley2007introduction}}]\label{lemma:covariance_bounds}
            Let $(\Omega,\mcF,\pr)$ be a probability space then for any two sub-$\sigma$-algebras $\mcA$ and $\mcB$ we have that for $X \in L^2(\mcA)$ and $Y\in L^2(\mcB)$ that 
                \[
                \abs{\cov(X,Y)} \leq \rho(\mcA,\mcB) \norm{X}_{L^2} \norm{Y}_{L^2} \ ,
                \]
                \[
                    \abs{\cov(X,Y)} \leq \oldpsi(\mcA,\mcB) \norm{X}_{L^1}\norm{Y}_{L^1} \ ,
                \]
            and
                \[
                    \abs{\cov(X,Y)} \leq 2\varphi(\mcA,\mcB)\norm{X}_{L^1}\norm{Y}_{L^\infty}\, .
                \]
        \end{lemma}

        Now, given a sequence of random variables $(X_n)_{n\in\mbN}$ sequences of mixing coefficients are constructed  as follows: 

    \begin{dfn}\label{def:mixing_coefficients}
        Let $(X_n)$ be a sequence of random variables and let $\mcF_k = \sigma(X_n : n \leq k)$ and $\mcF^k = \sigma(X_n : n \geq k )$. Then we define, 
            \begin{align}
                \rho_n \ &:= \ \sup_{k\in\mbN} \ \rho(\mcF_k, \mcF^{n+k})\, ,\label{rho}\\
                \oldpsi_n \ &:= \  \sup_{k\in\mbN} \ \oldpsi(\mcF_k, \mcF^{n+k})\, ,\label{oldpsi}\\
                \varphi_n \ &:= \ \sup_{k\in\mbN} \ \varphi(\mcF_k, \mcF^{n+k})\, . \label{phi}
            \end{align}
    \end{dfn}

    It is clear from the definition that $\oldphi_n, \oldpsi_n, \varphi_n$ are all non-increasing in $n$. 
    The convergence of either sequence of coefficients to $0$ displays an ``asymptotic independence" or an ``asymptotic decorrelation" of the sequence of random variables.  
    We say that the sequence $(X_n)_{n\in\mbN}$ is $\rho$-mixing if one has that  $\rho_n \to 0$ as $n\to\infty$. We also note that the mixing condition $\rho_n\to 0$ was introduced in \cite{kolmogorov1960strong}. 
    The sequence of random variables  $(X_n)_{n\in\mbN}$ is $\oldpsi$-mixing if one has that  $\oldpsi_n \to 0$ as $n\to\infty$. This $\oldpsi$-mixing condition was first introduced in \cite{philipp1969central}. Finally, the sequence $(X_n)_{n\in\mbN}$ is called $\varphi$-mixing if $\varphi_n \to 0$ and was introduced in \cite{ibragimov1959some}.
   
    The following observations are also useful.
    
    \begin{prop}\label{prop:rho_small_psi}
        Let $(\Omega,\mcF,\pr)$ be a probability space then for any two sub $\sigma$-algebras $\mcA$ and $\mcB$ we have that 
            \[\rho(\mcA,\mcB) \leq \oldpsi(\mcA,\mcB) \, .\]
    \end{prop}
       \begin{proof}
            It is enough to establish that for two random elements $f \in L^2(\mcA)$ and $g\in L^2(\mcB)$ with $\vari(f),\vari(g) \neq 0$ we have that $\corr(f,g) \leq \oldpsi(\mcA,\mcB)$. Since one has that $\corr(f-\mbE[f], g-\mbE[g]) = \corr(f,g)$ we may, without loss of generality, assume that both $f$ and $g$ are centered (i.e. $\mbE[f] = \mbE[g] = 0$). 
            Indeed, this is the case as 
                \begin{equation}
                    \abs{\corr(f,g)} 
                        = 
                            \dfrac{
                                \abs{\mbE[fg]}
                            }
                            {
                                \norm{f}_{L^2} \norm{g}_{L^2}
                            } 
                        \leq 
                            \dfrac{
                                \abs{\mbE[fg]}
                            }
                            {
                                \norm{f}_{L^1}\norm{g}_{L^1}
                            } 
                        \leq 
                            \oldpsi(\mcA,\mcB) \, .
                \end{equation}
            where we have used \Cref{lemma:covariance_bounds} together with the assumption that $\mbE[f] = \mbE[g] = 0$. 
        \end{proof}

    \begin{prop}[%
    {\cite[Lemma 7.1]{doob1942stochastic}} and \cite{bradley2005basic,peligrad83noteon}%
    ]
    \label{prop:phi_small_psi}
        Let $(\Omega,\mcF,\pr)$ be a probability space. Then for any two sub $\sigma$-algebras $\mcA$ and $\mcB$ we have that
            \[
                \rho(\mcA,\mcB) 
                    \leq 2 (\varphi(\mcA,\mcB))^{1/2}(\varphi(\mcB,\mcA))^{1/2}
                    \leq 2 \varphi(\mcA,\mcB)^{1/2}\, .
            \]
    \end{prop}

    \begin{remark}
        Due to \Cref{prop:rho_small_psi} and \Cref{prop:phi_small_psi}, we see that to prove  \Cref{thm:mixing}, one only needs to provide a proof for the case $\rho_n \to 0$. 
    \end{remark}

\subsection{Proof of \texorpdfstring{\Cref{thm:mixing}}{Theorem 5}}
Equipped with the definitions from the prior section, we can obtain \Cref{thm:mixing} using the mixing properties of the subfamily of the dynamical propagators $\{\phi{n-1}{n}\}_{n\in\mbN}$. 
    Similar to before we define $\mcF_k$ to be the forward $\sigma$-algebra generated by $\phi{n-1}{n}$ for all $n\leq k$ and $\mcF^k$ the backwards $\sigma$-algebra generated by $\phi{n-1}{n}$ for all $n\geq k$. 
    We first provide a proof of the following lemma that states that if the subsystem  $\{\phi{n-1}{n}\}_{n\in\mbN}$ is $\rho$-mixing, then the averages $\mbE_{\pr}[\cnum{\phi 0 n}]$ converge (to 0) superpolynomially (i.e. faster than any polynomial). 
    \begin{lemma}\label{lemma:C_decay_in_poly}
        For the subfamily of dynamical propagators $\{\phi {n} {n+1}\}_{n\in\mbN_0}$ if one has that $\rho_n \to 0$ then for any $p \in \mbN$ we have that there is some $C_p<\infty$ such that 
            \[
            \mbE_{\pr}[\cnum{\phi 0 n}] \leq \dfrac{C_p}{n^p} \, .
            \]
    \end{lemma}
        \begin{proof}
            First note that for any $t$, if $\cnum{\phi 0 t} \in \mcA$, for some $\sigma$-algebra $\mcA$ then it must be the case that $\cnum{\phi 0 t}\in L^2(\mcA)$ as $0\leq \cnum{\phi 0 t} \leq 1$ almost surely.
            Now let $r,s,t \in \mbN$ then we have that $\cnum{\phi {0}{t+s+r}} \leq\cnum{\phi 0 r}\cdot \cnum{\phi{r}{r+s}} \cdot \cnum{\phi{r+s}{r+s+t}}$. 
            But we have that $\cnum{\phi{r}{r+s}}$ is almost surely bounded above by $1$, therefore we have that $\cnum{\phi {0}{r+s+t}} \leq \cnum{\phi {0}{r}}\cdot\cnum{\phi{r+s}{r+s+t}}$. Hence, we obtain that
                \begin{align}
                    \avg{\cnum{\phi{0}{r+s+t}}}
                     &
                        \leq \avg{\cnum{\phi{0}{r}} \cdot \cnum{\phi{r+s}{r+s+t}}}\,.
                \end{align}

            Now we use \Cref{lemma:covariance_bounds} together with the fact that $\cnum{\phi{0}{r}}\in L^2(\mcF_{r})$ and $\cnum{\phi{r+s}{r+s+t}} \in L^2(\mcF^{r+s})$, yielding 
                \begin{equation}
                    \avg{\cnum{\phi{0}{r+s+t}}}
                        \leq \avg{\cnum{\phi 0 r}}\cdot \avg{\cnum{\phi {r+s} {r+s+t}}} 
                            + \rho_s \cdot
                                (\avg{(\cnum{\phi{0}{r}})^2})^{1/2}
                                \cdot
                                (\avg{(\cnum{\phi {r+s}{r+s+t}})^2})^{1/2}\, .
                \end{equation}
            But we have that $\pr$-almost surely, 
                \begin{equation}
                    \cnum{\phi {r+s} {r+s+t}}  = \cnum{\phi 0 t\circ \theta_{r+s}}\,,
                \end{equation}
            and that $\cnum{\phi 0 t}^2 \leq \cnum{\phi 0 t}$. 
            Therefore, we have that 
                \begin{equation}\label{lemma:C_decay_in_poly_ineq_1}
                    \avg{\cnum{\phi{0}{r+s+t}}}
                      \leq 
                        \avg{\cnum{\phi{0}{r}}}
                            \cdot
                            \avg{\cnum{\phi{0}{t}}} 
                        + 
                            \rho_s
                            \cdot
                            (\avg{\cnum{\phi{0}{r}}})^{1/2}
                            \cdot
                            (\avg{\cnum{\phi{0}{t}}})^{1/2} \, .
                \end{equation}
    
            For the remainder of the proof, we denote $c(n,m) := \avg{\cnum{\phi{n}{m}}}$ for all $n,m\in\mbN$ with $n\leq m$, for notational convenience. 
            As such, inequality \ref{lemma:C_decay_in_poly_ineq_1} above establishes that for all $r,s,t\in\mbN$ we have 
                \begin{equation}\label{lemma:C_decay_in_poly_ineq_2}
                    c(0, r+s+t) \leq c(0,r) \cdot c(0,t) + \rho_s\cdot c(0,r)^{(1/2)}\cdot c(0,t)^{1/2} \, .
                \end{equation}

            Now let $p\in\mbN$ be fixed ($p$ represents the order of the polynomial in which the \Cref{lemma:C_decay_in_poly} states the polynomial convergence rate is).
            Then using the convergence $\lim_{n\to\infty} c(0,n) \to 0$, in \Cref{lemma:expectation_of_cnum_to_0} and the assumption that $\rho_n\to0$, we have that there exists $M, N_0 \in \mbN$ with $M\leq N_0$ such that $\rho_M + c(0,N_0) \leq 1/4^p$. 
            Furthermore from \Cref{lemma:C_decay_in_poly_ineq_2} with $r=t = N_0$ and $s = M$ we have that 
            \[  
                c(0, 2N_0+M) \leq c(0,N_0)^2 + \rho_M c(0,N_0) 
            \]
          
            Now consider the sequence $\{N_i\}_{i\geq 0}\subseteq \mbN$ where     
                \[N_i = 2^iN_0  + (2^i-1)M \, .\]
            Using \ref{lemma:C_decay_in_poly_ineq_2} with $r=t = N_i$ and $s = M$ we are able to obtain that $\forall i\in\mbN\cup\{0\}$
                \[
                    c(0, N_{i+1}) = c(0, 2N_i+M) 
                    \leq c(0,N_i)^2 + \rho_M\cdot c(0,N_i) \leq \dfrac{1}{4^{p(i+1)}} c(0, N_0) \, .
                \]
            Now, given a $n\in\mbN$ that is larger than $N_0$ we must have that $N_i \leq n < N_{i+1}$ for some $i\in\mbN$. 
            This gives us that 
                \begin{equation}
                    c(0,n)\leq c(0,N_i) \leq \dfrac{c(0,N_0)}{4^{i\cdot p}} , 
                    \quad \forall n\geq N_0 \, .
                \end{equation}
            However as $M \leq N_0$ we also have from the construction of $N_i$ that $N_i \leq 2^{(i+1)}N_0 \leq 4^{i-1}N_0$ for all $i\geq 3$. 
            Therefore for large $n$ where $n\geq N_i$ with $i\geq 3$, since $n < N_{i+1}$ we have that $n < 4^{i}N_0$, giving us that
                \[
                    \left(\dfrac{N_0}{n}\right)^p \geq  \dfrac{1}{4^{p\cdot i}} \, .
                \]
            Thus, for sufficiently large $n$ we have that 
                \[
                    c(0,n) \leq c(0,N_0) \cdot (N_0)^p \cdot \dfrac{1}{n^p} \, .
                \]
            Since this holds for $n$ large enough, one can then find some constant $C_p$  such that 
                \[
                    c(0,n) \leq C_p\cdot \dfrac{1}{n^p}
                \]
            for all $n\in\mbN$ providing the claim in \Cref{lemma:C_decay_in_poly}.     
        \end{proof}
        
    We now show that under an additional independence assumption, the expectations $\mathbb{E}[\cnum{\phi_{0}^{n}}]$ decay exponentially fast.
    
    \begin{lemma}\label{lemma:expectation_discrete}
        If the discrete family $(\phi{n-1}{n})_{n\in\mbZ}$ is jointly independent then we have that
            \[
                 \avg{\cnum{\phi 0 n}}  \leq   C (e^{ -\delta n}) \ ,
            \]
        for some $\delta>0$ and $C>0$.
    \end{lemma}
        \begin{proof}
            Since we know from \Cref{lemma:expectation_of_cnum_to_0} that $\avg{\cnum{\phi 0 n}} \to 0$ we must have that there exists some $N_0\in\mbN$ such that $\avg{\cnum{\phi 0 n}} \leq 1/2$ for all $n\geq N_0$. 
            Now let $\delta>0$ with $\delta < \ln(2)/(2N_0)$.
            Now for given $n\in \mbN$, with $n \geq 2N_0$, we can find $i\in\mbN$ such that $n \in [2iN_0, 2(i+1)N_0)$ so that
                \[
                    \avg{\cnum{\phi 0 n}} \leq \avg{\cnum{\phi{0}{2iN_0}}} \leq \left(\avg{\cnum{\phi{0}{N_0}}}\right)^{2i} \, .
                \]
            But we also have that $\avg{\cnum{\phi{0}{N_0}}} \leq 1/2$ and that $\delta < \ln(2)/(2N_0)$. Thus, $1/2 \leq e^{-2N_0\delta}$ giving us 
                \[
                    \avg{\cnum{\phi 0 n}} \leq e^{-4 i \delta N_0} \leq e^{-n\delta}\, ,
                \]
            where the last inequality uses $n \leq 2(i+1)N_0 <  4iN_0$. Since this holds for $n$ large enough, we can thus find a constant $C$ so that we have that $\avg{\cnum{\phi 0 n}} \leq C e^{-n\delta}$ for all $n\in\mbN$. 
        \end{proof}

\fourthresult*
    \begin{proof}
        From the proof of \Cref{thm:existence_of_z_0} and \Cref{thm:super-op-convergence} we have the following inequalities:
            \begin{enumerate}
                \item $\norm{\phiadj s t \proj \updelta-Z'_s} \leq 2\cnum{\phi s t}$.
                \item $\norm{\phi s t  \proj \updelta - Z_t} \leq 2\cnum{\phi s t} $.
                \item $\norm{\dfrac{\phi s t(\updelta)}{\tr{\phiadj s t}(\mbI)} - \opp{s}{t}(\updelta)}\leq 8\cnum{\phi s t}$.
            \end{enumerate}
        for all $\updelta\in\states$.  By \Cref{prop:rho_small_psi} it is enough to establish the result when $\rho_n\to 0$, but then we also have, from \Cref{lemma:C_decay_in_poly}, that there is some $C_p$ such that 
            \begin{alignat}{2}
                \int_\Omega{\cnum{\phi s t}} \, d\pr(\omega) 
                    &= \int_\Omega{\cnum{\phi{0}{t-s{}}}} \, d\pr(\omega)
                        &&= \avg{\cnum{\phi {0}{t-s}}}\\
                    &\leq \avg{\cnum{\phi{0}{\floor{t-s}}}}
                        &&\leq C_p \dfrac{1}{\floor{t-s}^p}\\
                    &   
                        &&\leq 2^p C_p \dfrac{1}{(t-s)^p}\, .
            \end{alignat}
        Where the last inequality is for $t-s>1$, and we have used \Cref{lemma:C_decay_in_poly}. This proves the first half of \Cref{thm:mixing}. For the case that the family of dynamical propagators has independent increments, then we apply \Cref{lemma:expectation_discrete} and the result follows by the same argument. 
    \end{proof}

Finally, we recall \Cref{thm:large_probability}, proof of which follows directly from the inequalities established in the proof of \Cref{thm:mixing} and Markov's inequality. 
\probresult*
\begin{proof}
    Recall from the proof of \Cref{thm:mixing} we have that for $s,t$ with $t-s>1$ we have with probability $1$ that, 
    \begin{equation}
        \norm{\dfrac{\phi s t}{\tr{\phiadj s t}(\mbI)} - \opp{s}{t}}\leq 8\cnum{\phi s t}\, .
    \end{equation}
    But we readily have from Markov's inequality that for any $a>0$, 
    \begin{equation}
        \pr\left\{\cnum{\phi s t} > a \right\} \leq \dfrac{1}{a}\avg{\cnum{\phi s t}} = \dfrac{1}{a}\avg{\cnum{\phi {0}{t-s}}}\, .
    \end{equation}
    Now given $p \in \mbN$ from \Cref{lemma:C_decay_in_poly} and the proof above, we have that there is a constant $C_{2p}$ such that whenever $t-s>1$ we have 
    \begin{equation}
        \pr\left\{\cnum{\phi s t} > a \right\} \leq \dfrac{1}{a}\avg{\cnum{\phi s t}}
        \leq \dfrac{2^{2p}C_{2p}}{a} (t-s)^{-2p}
        \, .
    \end{equation}
    Therefore, we have that 
    \begin{equation}
        \pr\left\{ \norm{\dfrac{\phi s t}{\tr{\phiadj s t}(\mbI)} - \opp{s}{t}} \leq a\right\} 
        \geq 
        \pr \left\{ 8 \, \cnum{\phi s t } \leq a\right\} 
        \geq1 - \dfrac{8\cdot2^{2p}C_{2p}}{a} (t-s)^{-2p}\, .
    \end{equation}
    The first part of the proof is concluded by taking $a = (t-s)^{-p}$, and the proof for the case of independent increments is similar and uses \Cref{lemma:expectation_discrete} instead of \Cref{lemma:C_decay_in_poly}. 
    Similar result holds for the quantity $\pr\{\norm{\phi s t \cdot \updelta - Z_t} \le (t-s)^{-p}\}$. 
\end{proof}